\documentclass[referee,usenatbib]{mn2e}
\usepackage{epsfig}
\usepackage{subfigure}
\usepackage{color}
\newcommand{\kpc}{\>{\rm kpc}}
\newcommand{\Msun}{\>{\rm M_{\odot}}}
\newcommand{\msun}{\>{\rm M_{\odot}}}

\newcommand{\beq}{\begin{equation}}
\newcommand{\eeq}{\end{equation}}
\newcommand{\kpch}{\>{h^{-1}{\rm kpc}}}
\newcommand{\hmpc}{\>h^{-1}{\rm {Mpc}}}

\def\Msun{\,\rm M_{\odot}}

\newcommand{\mvir}{m_{\rm vir}}
\newcommand{\rvir}{r_{\rm vir}}

\newcommand{\rcool}{r_{\rm cool}}

\newcommand{\mcooldot}{\dot{m}_{\rm cool}}

\newdimen\hssize
\newdimen\hdsize 
\usepackage{color}

\usepackage[normalem]{ulem}

\begin{document}

%%%%%%%%%%%%%%%%%%%%%%%%%%%%%%%%%%%%%%%%%%%%%%%%%%%%%%%%%%%%%%%%%%%%%%%%%%
\title[Radiative Cooling in SAMs]
      {On the algorithms of radiative cooling in semi-analytic models}
\author[]
       {
                Yu Lu$^{1,2}$\thanks{E-mail: luyu@stanford.edu},
                Du\v{s}an Kere\v{s}$^{3,4,\dag}$,
                 Neal Katz$^{1}$,
                H.J. Mo$^{1}$,
        Mark Fardal$^{1}$,
                \and Martin D. Weinberg$^{1}$
\\
        $^1$ Department of Astronomy, University of Massachusetts,
        Amherst MA 01003-9305
\\
		$^2$ Kavli Institute for Particle Astrophysics and Cosmology, Stanford, CA 94309
\\
        $^3$ Institute for Theory and Computation, Harvard-Smithsonian Center for Astrophysics, Cambridge, MA 03138
\\
		$^4$ Department of Astronomy and Theoretical Astrophysics Center, University of California, Berkeley, CA 94720-3411, USA
\\
		$^{\dag}$ Hubble Fellow
                }

%%%%%%%%%%%%%%%%%%%%%%%%%%%%%%%%%%%%%%%%%%%%%%%%%%%%%%%%%%%%%%%%%%%%%%%%%%

\date{}

\pagerange{\pageref{firstpage}--\pageref{lastpage}}
\pubyear{2010}

\maketitle

\label{firstpage}

%%%%%%%%%%%%%%%%%%%%%%%%%%%%%%%%%%%%%%%%%%%%%%%%%%%%%%%%%%%%%%%%%%%%%%%%%%
\begin{abstract}
We study the behaviour of multiple radiative cooling algorithms implemented
in seven Semi-Analytic Models (SAMs) of galaxy formation, including a new model
we propose in this paper. We use versions of the models
without feedback and apply them to dark matter haloes growing in a cosmological
context, which have final virial masses that range from $10^{11}\msun$ to
$10^{14}\msun$. First, using simplified smoothly-growing halo models, 
we demonstrate that the different algorithms predict
cooling rates and final cold gas masses that differ by a factor of $\sim$5
for massive haloes ($\geq10^{12}\msun$).  The algorithms are in better
agreement for less massive haloes because they cool efficiently and, therefore,
their cooling rates are largely limited by the halo accretion rate.
However, for less massive haloes, all the SAMs predict less cooling than
corresponding 1D hydrodynamic models. Second, we study the gas accretion history 
of the central galaxies of dark matter haloes using merger trees. The inclusion of mergers alters 
the cooling history of haloes by locking up gas in galaxies within small haloes 
at early times. For realistic halo models, the dispersion in the cold gas 
mass predicted by the algorithms is 0.5 dex for high mass haloes and 
0.1 dex for low mass haloes, while the dispersion in the accretion rate is 
about two times larger. Comparing to cosmological SPH simulations, we 
find that most SAMs systematically under-predict the gas accretion 
rates for low-mass haloes but over-predict the gas accretion rates for 
massive haloes. Although the models all include both ``rapid'' and ``slow'' 
mode accretion, the transition between the two accretion 
modes varies between models and also differs from the simulations. 
Finally, we construct a new phenomenological model that explicitly incorporates cold halo
gas and a gradual transition between the cold and hot modes of gas accretion 
to illustrate that such a class of models can better match the results
from cosmological hydrodynamic simulations. The large dispersion in 
cooling rates between different SAMs influences parameter 
choices for other galaxy physics including star formation and feedback. 
Therefore, careful parameterisations of the multimode gas cooling and
accretion mechanisms in simulations are necessary to 
ensure that the predictions from SAMs are reliable.  
\end{abstract}
%%%%%%%%%%%%%%%%%%%%%%%%%%%%%%%%%%%%%%%%%%%%%%%%%%%%%%%%%%%%%%%%%%%%%%%%%%


%%%%%%%%%%%%%%%%%%%%%%%%%%%%%%%%%%%%%%%%%%%%%%%%%%%%%%%%%%%%%%%%%%%%%%%%%%
%\LARGE
\section{Introduction}\label{sec:intro}

The formation and evolution of galaxies has proven a challenging problem,
as it involves a range of complicated physical processes. Pioneering work 
based on simplified models (e.g. \citealt{White1978, White1991}) indicates
that at least radiative cooling, star formation, and feedback must
be included in an even minimally realistic galaxy formation model.
To make progress in understanding galaxy formation, two alternative
approaches have been pursued. One approach is to simulate directly the
relevant physical processes in a cosmological context
\citep[e.g.][]{Katz1991, Katz1992, Springel2003, Kerevs2005, Kerevs2009}. 
Although this might seem the ideal approach, it is limited by 
its computational costs and its ability to only resolve a relatively small 
dynamical range, which makes it necessary to include many important
``sub-grid'' physical processes, e.g. star formation, using heuristic
formulas. The second approach is the so-called ``semi-analytic'' models 
(SAMs, hereafter) \citep{White1991, Kauffmann1993, Somerville1999a, Cole2000}. 
These models use simple parameterisations for baryonic processes within 
a model of dark matter halo formation, using either actual dark matter 
simulations or a Monte-Carlo method for the dark haloes. The results from 
SAMs can be directly compared with a wide range of observations and, 
because SAMs are much less computationally expensive than direct simulations, 
they have been widely adopted for galaxy formation problems.  However, 
SAMs suffer from their own shortcomings. SAMs, by definition, are based 
on simplified models of physical processes.  The simplifications are
most often justified 
by the need to get a ``good fit'' to the observations, rather 
than a careful examination of their validity.
In addition, very different parameterisations have been adopted
for the same physical process by different research groups, and the
consequences of choosing one implementation over another have not been
carefully addressed.

The addition of radiative cooling is the first step in creating a SAM from
a dark-matter halo model.
Since all processes are coupled in galaxy formation, the
predicted amount of gas cooling affects the modelling of all
other processes.  Compared to star formation, feedback, and
other relevant processes, the fundamental physics of radiative cooling
is quite well understood, but the cooling rate within a halo depends
on the physical state of the gas throughout the halo
and is, therefore, still highly uncertain. 
Since the gas cooling rate is not directly observable in real galaxies,
the performance of a cooling model depends indirectly 
on models for the star formation, feedback,
etc. Numerical simulations including
gas dynamics can provide direct comparisons with cooling models in SAMs.
\citet{Benson2001c} compared the statistical properties of cooled gas
predicted in a cosmological hydrodynamic  simulation with that of 
a ``stripped-down'' SAM, which includes no more baryonic processes 
other than radiative cooling. They found that 
the global fractions of hot gas, cold gas, and
uncollapsed gas were consistent within 25\%, and the mass of gas
in the cold phase in the most massive haloes differed by no more than 50\%.
In a similar spirit, \citet{Yoshida2002}, \citet{Helly2003}, and \citet{Cattaneo2007}
performed similar comparisons in more detail. They extracted merger trees
from cosmological hydrodynamic simulations and made use of these same merger
trees in their SAMs. By doing so, they were able to compare the predictions
from those two approaches on an object-by-object basis.  They reached
a similar conclusion: the two methods, direct hydrodynamic simulations
and SAMs, predict roughly the same ``galaxy'' populations.
However, a recent SAM -- SPH simulation comparison by \citet{Saro2010} 
focused on BCG galaxies and found that although the two methods predict similar 
statistical properties of the galaxy population in general, BCGs in SPH 
simulations tend to have stronger starbursts at earlier times and less star 
formation at later times than their counterparts in SAMs. 
To isolate the effects of radiative cooling from the complex process of halo
formation,
\citet{Viola2008} compared their SAM cooling model to hydrodynamic simulations
of halo gas in hydrostatic equilibrium within isolated
dark matter haloes. 
The authors found that the MORGANA model could predict the cooling 
histories of the static halos with different masses in a remarkable 
agreement with their simulations.
Similarly, \cite{Keres2007} showed that, once the
proper profile of the gas is taken into account, the Bertschinger 
solution for the gas cooling in a static potential is well reproduced in Gadget-2 simulations of isolated halos.

Hence, it seems that studies by different groups have converged to the same
conclusion: SAMs can be tuned to agree with cosmological hydrodynamic
simulations.  However, there are still a number of outstanding issues
that have not yet been addressed. 
First, most of the comparisons mentioned limited their investigation to their own SAM 
or only included a couple of different models; 
a cross-check over a larger number of existing SAMs needs to be performed.
Indeed, a recent comparison between three SAMs by \citet{DeLucia2010} 
revealed that the discrepancy between the cooling rates of massive halos 
predicted by different models can be up to one order of magnitude. 
Second,
the cooling in higher mass haloes suffered from unphysical, numerically
enhanced cooling \citep[see][]{Springel2002, Keres2007}
in some of the hydrodynamic simulations.
Clearly one must use a simulation that avoids excess cooling
to compare with the SAM predictions. Third, the way the cooling models work
in detail needs to be understood. Most of the cooling models
are based on the same central idea \citep{White1991}: a one-dimensional self-similar
analytic solution of gas cooling in a static potential \citep{Bertschinger1989}.
However, in practise the various implementations are different, and one should
understand the implications of these differences before adopting a
particular cooling model. Fourth, the previous comparisons
with simulations usually focus on the final galaxy masses \citep[e.g.][]{Yoshida2002}.
But, since cooling is the process that fuels star formation, differences
in the cooling rates can affect the observed evolution of galaxies from high-$z$ to
the present time.  Hence, an understanding of the differences between
the models requires a comparison of cooling rates over a 
cosmological timescale. 
Fifth, even though the evolution of the cooling rate has been studied by some
authors \citep{Keres2007, Viola2008}, these studies assumed idealised initial
conditions that ignored the hierarchical nature of dark matter halo formation.
Because galaxies grow through a combination of cooling and hierarchical
merging, a study of the performance of cooling models in haloes growing
in a cosmological context is more relevant. 
Recently, \citet{Stringer2010a} demonstrated that the discrepancy between 
the predicted cold baryon mass in SAMs and simulations owes to the initial 
conditions and physical assumptions, not the choice of modelling technique. 

Finally, SPH simulations have 
shown that a large fraction of the baryonic mass in galaxies of all masses 
is acquired through ``cold-mode'' accretion of gas that is never shock heated 
to its halo virial temperature
\citep{Katz2003, Kerevs2005, Keres2007, Ocvirk2008, Kerevs2009, Brooks2009}. Similarly, analytic
arguments and 1D models \citep{Birnboim2003} showed that gas collapsing into
haloes will not be shock heated during its infall into low-mass haloes
\citep[see also][]{Binney1977}. However, except for a few attempts with
simplified models \citep{Cook2009, Khochfar2009, Kang2010, Benson2010b}, 
cold-mode accretion has not be implemented in SAMs.
Physical processes proceed on multiple timescales in SAMs.  In the
standard prescription, gas in
low mass haloes cools rapidly so that all of the infalling gas
is accreted on a short timescale. Filamentary, non-spherical infall of gas
in cold-mode accretion corresponds to a different physical picture of
gas infall than a rapid cooling of the virialised halo gas.
Furthermore, in simulations cold and hot-mode accretion can co-exist
(unlike ``rapid'' and ``slow'' cooling regimes in SAMs).  Therefore, cold-mode
accretion can be important even in massive haloes dominated by hot,
virialised gas. However, cold-mode accretion
will also be limited by the infall of baryons into haloes.  One could
argue that these physical details are not needed to model the 
cold-mode accretion in low mass haloes, and that the ``rapid''
cooling regime in SAMs is a sufficient proxy.
Since different SAMs define the transition between the two modes in
different ways, a detailed comparison between the models and
simulations is needed to see if the gas accretion rates are
captured realistically in both regimes.

Motivated by these open issues, we investigate the behaviour of six
representative cooling algorithms commonly adopted by SAMs
that are widely cited in the recent literature, namely those
of \citet{Somerville1999a}, \citet{Cole2000}, \citet{Hatton2003}, \citet{Kang2005},
\citet{Croton2006a} and \citet{Monaco2007}. We implement ``stripped-down'' versions
of these models by ``turning off'' all the baryonic processes other
than radiative cooling to make predictions of the cooling histories
for different dark matter haloes.  
We then compare the predictions of 
these models to each other and to those of hydrodynamic simulations.
In particular, we compare the predicted cooled gas accretion rates 
and the cumulative cold gas masses of the central objects of dark matter 
haloes with a wide range of virial masses as a function of redshift.
Our goal is to present a detailed comparison between the different 
cooling algorithms commonly used in SAMs and between the predictions 
made by these SAMs and hydrodynamic simulations.  Discrepancies 
between different SAM algorithms and between the SAMs and the simulations 
will help guide the improved modelling of radiative cooling processes 
in galaxy formation.
 
The plan of the paper is as follows.  In \S\ref{sec:model}, we
review the algorithms of radiative cooling adopted by different
groups and describe in detail how the algorithms are implemented in
SAMs. In \S\ref{sec:res}, we compare the results of these models and 
of simulations in situations of increasing complexity. As a first step,
we apply the algorithms to an isolated halo model (\S\ref{sec:res}.1) with
temporally smooth, spherically-averaged accretion.
Thus all the cooling occurs only
in the main branch of the halo merger tree but not in any of the smaller mass
progenitors. Using this model, we are able to filter out the complications
introduced by halo merger trees and focus our study on the behaviour
of the cooling algorithms themselves. We use a spherically
symmetric hydrodynamic code \citep{Lu2007} to simulate the cooling in these
accreting dark matter haloes to compare with the predictions of the SAM 
cooling algorithms. Next,
we apply the stripped-down models to full merger trees of dark matter haloes
in which we allow cooling in progenitors (\S\ref{sec:res}.2).
We repeat the predictions made in the first step and study how these 
cooling algorithms behave in realistic merging haloes. We then apply the 
stripped-down SAMs to a cosmological volume, comparing the different SAMs 
with each other and with the cosmological hydrodynamic simulation of 
\citet{Kerevs2009} (\S\ref{sec:res}.3). We find that none of the models 
compares well with the simulated cold-mode and hot-mode accretion 
rates across all halo masses. To address this problem,
we propose a new model that explicitly incorporates 
``cold'' accretion (\S\ref{sec:new}). We then compare the predictions of the new model 
with the simulation and other SAMs. In \S\ref{sec:dis}, we summarise our 
results from the comparisons and discuss possible steps one might take to 
improve models of radiative cooling for galaxy formation.

We assume a cosmology with $\Omega_{\rm m}=0.26$, $\Omega_{\rm \Lambda}=0.74$, $\sigma_8=0.75$, $h=0.71$,
and $\Omega_{\rm b,0}=0.044$. To focus our investigation on the algorithms themselves, we
assume that the gas has zero metallicity throughout
its entire evolution.  We interpolate the tabulated cooling
function of \citet{Sutherland1993} for all the cooling rates used in our calculations
except in the SPH simulation, which uses the cooling function from \cite{Katz1996}.
We set the ionising UV background in our SAMs to zero except when testing
against the SPH simulation, which does contain a UV background; the
differences between these two cases are small for the halo mass range
considered.

\section{Cooling Algorithms}\label{sec:model}
All the cooling models in SAMs share some parts in common because they
are derived from \citet{White1991}.
However, the cooling algorithms differ in detail.
In this section, we review the prescriptions of 
the six different cooling models in detail.

\subsection{Common Features}

The cooling recipes are designed to capture the most important aspects of galaxy formation, even though
cooling in the real Universe involves many complications, such as clumpy 
gas distribution, complex chemical distributions, a multi-phase medium etc.
In SAMs, one assumes that the gas shock-heats to the host halo's virial temperature 
with a simply-described smooth density distribution.
In addition, one assumes that the chemical abundances
are well mixed. Since the hot gas is in thermal and
ionisation equilibrium, the cooling rate simply depends on the temperature,
the density, and the metallicity of the gas. The primary cooling processes
relevant to galaxy formation are
(i) collisional excitation and ionisation, which are important for haloes with
intermediate virial temperatures ($10^4<T<10^6{\rm K}$),
(ii) recombination, and (iii) bremsstrahlung radiation, which dominates 
at $T\sim10^7{\rm K}$ \citep{Thoul1995}. As these radiative processes occur, 
the hot halo gas loses its thermal energy and hence its pressure support and, 
conserving angular momentum, eventually collapses onto a galaxy disk in 
the centre of the halo \citep[e.g.][]{Mo1998}. 
Since radiative cooling depends on the square of 
the density, the central hot gas, which is denser, cools faster than the gas 
in the outer halo.  This inside-out cooling is a common feature in SAMs.

The general framework for gas cooling in SAMs is as follows.
For a given dark matter halo at time $t_0$, the gas is distributed
in the halo following an assumed density profile. In most models, the
density distribution is a singular isothermal density profile,
\begin{equation}\label{equ:isoden}
\rho_{\rm gas}(r)={m_{\rm gas0} \over 4\pi r_{\rm vir} r^2},
\end{equation}
where $r_{\rm vir}$ is the virial radius of the halo and
$m_{\rm gas0}$ is the hot gas mass contained within the halo.
Some models add a constant density core to this profile and include
a time-varying core radius \citep{Mo1996, Cole2000, Bower2001, Benson2003c}.
In this paper, we mainly focus on the singular isothermal density profile,
but we include two models with cores (the GalICS model and one variant of the Cole model).  
Using these two models, we can gauge the effects of a cored gas profile 
on the cooling rates. 

The models assume that the halo gas shock-heats uniformly
to the virial temperature of the halo,
\begin{equation}\label{equ:temp}
T_{\rm gas}(r)=T_{\rm vir}= 35.9\left(\frac{V_{\rm c}}{\rm km s^{-1}}\right)^2{\rm K},
\end{equation}
where $V_{\rm c}$ is the circular velocity of the halo at the virial radius $\rvir$.
To determine the radius within which the gas is able to cool, one calculates
the cooling time of the gas as a function of the radius from the halo centre, $r$,
\begin{equation}\label{equ:cooltime}
\tau_{\rm cool}(r)=\frac{3}{2}{\mu m_{\rm H} kT_{gas}
\over \rho_{gas}(r)\Lambda(T_{gas},Z_{gas})},
\end{equation}
where $\mu$ is the mean molecular weight, $m_{\rm H}$ is the mass of the hydrogen atom,
$\rho_{\rm gas}$ and $T_{\rm gas}$ are the gas density and temperature at that radius,
$Z_{\rm gas}$ is the mean metallicity of the gas, and $\Lambda$ is the
cooling function. 
We tabulate $\Lambda$ for the cooling processes over a range of temperature and
metallicity \citep{Sutherland1993, Katz1996}. 
To simplify the physics involved and study the simplest cooling case, 
in this work we do not follow chemical enrichment and, therefore, 
we only use the cooling rates corresponding to `primordial' gas composition. 
One defines the halo cooling radius by equating 
the cooling time to a pre-defined timescale.  In practise, this implies that
the hot gas inside of cooling radius can cool within the pre-defined timescale.
The models also distinguish between a period of ``rapid'' cooling, 
typically when the cooling radius is larger than the virial radius, which occurs in
low mass haloes, and a period of ``slow'' cooling,  typically when the cooling 
radius is smaller than the virial radius, which occurs in massive haloes.
This separation into two 
cooling regimes is reminiscent of the two modes of hot and cold accretion 
seen in simulations \citep{Kerevs2005, Kerevs2009}.

We now briefly review how 
the cooling models are implemented in each of the six models. For a complete and 
detailed description of the physics included in each model, including 
the radiative cooling, see \citet{Croton2006a} for the
Croton model (also known as the Munich model), \citet{Kang2005} for the Kang model,
\citet{Cole2000} for the Cole model (also known as the Durham model or GALFORM model),
\citet{Hatton2003} for the GalICS model, \citet{Somerville1999a} for the Somerville model, 
and \citet{Monaco2007} for the MORGANA model.

\subsection{The Croton Model}\label{subsec:croton}
The halo merger trees are stored at a series of discrete times
from high redshift ($z\sim20$) to the present time. 
The model further subdivides every time interval between two 
snapshots into a number of ($\sim10$) fine time 
steps to allow for gas phase exchange (converting from the hot phase
to the cold phase) and for the commensurate evolution of galaxy properties.
At every model time step, the halo gas distribution is reset
to a singular isothermal profile.  
The total hot gas mass, $m_{\rm gas0}$, is 
the total baryonic mass in the halo minus the baryonic mass that has already 
been converted into cold gas or stars and any gas that has been  ejected from 
the halo by feedback.   This leads to the expression
\begin{equation}\label{equ:gasmass}
m_{\rm gas0}=f_{\rm b} m_{\rm vir} -
\sum_i \left[m_*^i + m_{\rm cold}^i + m_{\rm eject}^i\right],
\end{equation}
where $f_{\rm b}=\Omega_{\rm b, 0}/\Omega_{\rm m}$ is the universal baryon
fraction, and $m_*$,$m_{\rm cold}$, and $m_{\rm eject}$
are the masses in stars, cold gas, and ejected gas, respectively.
The sum is over all the galaxies (central and satellites) in the halo.
In our stripped-down model, we ignore both star formation and feedback, which
simplifies equation (\ref{equ:gasmass}). The hot gas temperature is reset 
to the virial temperature of the current halo using equation
(\ref{equ:temp}). The Croton model defines the cooling radius by
equating 
$\tau_{\rm cool}$ (eq. \ref{equ:cooltime}) to the halo dynamical time, $\tau_{\rm dyn}=r_{\rm vir}/V_{\rm c}$.
If the cooling radius is smaller than the virial radius, the cooling is 
quiescent or ``slow'' and the cold gas accretion rate is the instantaneous 
hot gas mass flux through the cooling radius:
\begin{equation}
\mcooldot=4\pi \rho_{\rm gas}(\rcool)\rcool^2
{{\rm d} \rcool \over {\rm d}t}=0.5m_{\rm gas0}{r_{\rm cool} \over r_{\rm vir} \tau_{\rm cool}} 
= 0.5m_{\rm gas0}{r_{\rm cool}V_{\rm c} \over r_{\rm vir}^2}.
\end{equation}
If the cooling radius is larger than the virial radius, the cooling is
in the catastrophic (``rapid'' cooling) regime and it is assumed that
all the hot gas cools onto the central galaxy instantaneously. In this
regime the cooling rate is effectively the total hot gas mass divided
by the time step $\Delta t$ \footnote{ The reference paper did not
  precisely describe the prescription for the ``rapid'' mode accretion rate. We
  translate their prescription of immediate cooling into equation
  (\ref{equ:rapidcool}). Also see \citet{DeLucia2010}.  }
\begin{equation}\label{equ:rapidcool}
\mcooldot={m_{\rm gas0} \over \Delta t}.
\end{equation}
If the halo stays in the ``rapid'' mode, the central galaxy
accretion rate equals the halo baryon accretion rate. However, 
if the halo accretion becomes rapid later and contains some residual hot gas, 
the galaxy accretion rate can be higher than the halo accretion rate.

\subsection{The Kang Model}\label{subsec:kang}
This model is similar to the Croton model but derives the cooling
radius by equating the cooling
time $\tau_{cool}$ to the
the Hubble time, $\tau_{\rm H}$, to evaluate the cooling radius at each
time step and uses $\tau_{\rm H}$ to derive the accretion rates
in both the ``slow'' and the ``rapid'' cooling regimes.  In the ``slow'' 
cooling regime, when $r_{\rm cool}<\rvir$, the cooling rate is
\begin{equation}
\mcooldot=0.5m_{\rm gas0}{r_{\rm cool} \over \tau_{\rm H} r_{\rm vir}};
\end{equation}
and in the ``rapid'' cooling regime,
\begin{equation}
\mcooldot={m_{\rm gas0} \over \tau_{\rm H}}.
\end{equation}
There is a factor of two jump in the cooling rate for haloes at the
transition between these two regimes.
If one defines 
the virial radius to be the radius that contains a mean density that is 200 
times the critical density, for an isothermal density profile the cooling 
radius in the Croton model is $\sqrt{10}$ times smaller than in the Kang
model since the halo dynamical time is $1/10$ of the Hubble time.
This makes the cooling rates in the Croton model about 3 times higher 
in the ``slow'' regime than in the Kang model at a fixed halo gas mass.

\subsection{The GalICS Model}\label{subsec:galics}

As in the Croton and Kang models, the GalICS model redistributes
the hot gas in the halo at every time step. However, this
model assumes a different profile than the previous two models: 
an isothermal sphere with a fixed core radius of 0.1 kpc 
\citep{Hatton2003, Cattaneo2007}, i.e.
\begin{equation}\label{equ:core}
\rho_{\rm gas}(r)=\rho_0 \rvir^2/(r^2+r_{\rm core}^2),
\end{equation}
where
\begin{equation}
\rho_0={m_{\rm gas0} \over 4 \pi \rvir^2 \left[\rvir-r_{\rm core} \arctan(\rvir/r_{\rm core})\right]}.
\end{equation}
Since the core size is small, the core only affects small haloes.
At each time step, the model assumes that hot gas can cool and collapse to
fuel the central galaxy if both the cooling time $\tau_{\rm cool}$
[See Eq.(\ref{equ:cooltime})] and the free-fall time $\tau_{\rm ff}=r/V_{\rm c}$
of the gas at radius $r$ are shorter than the time step $\Delta t$:
\begin{equation}\label{equ:coolrate3}
\mcooldot={m_{\rm gas}(r_{\rm min}) \over \Delta t} = {4\pi \rho_0 \rvir^2 \left[r_{\rm min} - r_{\rm core}\arctan(r_{\rm min}/r_{\rm core})\right] \over \Delta t},
\end{equation}
where $r_{\rm min}=\min[r_{\rm cool}, r_{\rm ff}, \rvir]$ at the current time step.
If the free-fall radius $r_{\rm ff}$ or the virial radius determines the cooling rate, 
then the halo gas cools in the ``rapid'' regime. If the cooling radius
is smaller than the free-fall radius and the virial radius, then the cooling radius determines
the cooling rate and the halo is in the ``slow'' cooling regime.
The radius $r_{\rm min}$ depends on the time step and,
therefore, the corresponding cooling rate and cooling regime also depends on 
the time step. Recall that $\tau_{\rm cool} \propto 1/\rho_{\rm gas}$ and
$\tau_{\rm ff} \propto r \propto 1/\sqrt{\rho_{\rm DM}}$,
which means that for small $\Delta t$ a halo is always in the free fall regime,
because the cooling time is always shorter than a free fall time in the centre, 
as long as $r_{\rm cool}$ is larger than the core radius.
If the time step is such that $r_{\rm ff}$ is a sizable fraction of $r_{\rm vir}$,
a halo then enters the ``slow'' cooling dominated regime. The real situation is more
complicated as the transition also depends on the total gas mass in the halo. 
For this reason, caution should be taken when we study the prediction for the
``slow'' cooling regime of this model.

\subsection{The Cole Model}\label{subsec:cole}

The Cole model uses the halo merger trees differently than in the
previous models. It divides
the growth history of the halo into a series of generations.
The initial redshift, the top of a tree, defines the first generation
of the halo.  A new generation begins when the halo doubles its virial mass;
this time is defined as a new birth time ($t_{\rm form}$).  
The gas density profile of a halo does not reset at every time step as in
the previous models but only at every birth time. In other words, the
halo
mass doubles at each generation.
The gas density profile outside
the cooling radius only changes after each generation.
At each time step, the model calculates the cooling radius, $r_{\rm cool}$,
where $\tau_{\rm cool}=t_{\rm age}$ and $t_{\rm age}$ is the lifetime of
the halo, i.e. $t_{\rm age}=t-t_{\rm form}$. At the same time, the model also
calculates the free-fall radius, $r_{\rm ff}=t_{\rm age}V_{\rm c}$.
The model assumes that only the hot halo gas within both of these two radii is
able to cool and accrete onto the central galaxy. Hence, the effective
cooling radius is $r_{\rm min}=\min[r_{\rm cool}, r_{\rm ff}, \rvir]$
at the current time step.
This model adopts a core density profile described by equation (\ref{equ:core}).
The more sophisticated models derived from this prescription allow
the core size and the density slope to evolve as a result of cooling and feedback
energy injection \citep{Bower2001, Benson2003c, Bower2006}.
To illustrate the behaviour of this model, we choose the core radius to
be  $r_{\rm core}=0.1\rvir$.
The cooling rate in a time step, $\dot{m}_{\rm cool}$,
equals the mass of the hot gas enclosed in the spherical shell between
the $r_{\rm min}$ of the previous time step
and that of the current time step divided by the time step, $\Delta
t$. This yields
\begin{equation}\label{equ:cr_cole}
\mcooldot={m_{\rm gas}(r_{\rm min}) -m_{\rm gas}'(r'_{\rm min}) \over \Delta t},
\end{equation}
where $r'_{\rm min}$ is the effective cooling radius of the last time step, and
$m_{\rm gas}(r)$ is the gas mass enclosed by radius $r$:
\begin{equation}
m_{\rm gas}(r) = 4 \pi \rho_{\rm 0} \rvir^2 \left[r-r_{\rm core}\arctan(r/r_{\rm core})\right].
\end{equation}
If the effective cooling radius  $r_{\rm min}$ equals the cooling
radius, cooling is in the ``slow'' regime; if it equals the free-fall
radius or the virial radius, cooling is in the ``rapid'' regime.
The cooling rate is the same as in the GalICS
model (eq.\ref{equ:coolrate3}) at the onset of a new halo generation
for the same value of $r_{\rm core}$. However, since the gas configuration
does not change between halo generations in the Cole model, but does change in the GalICS
model, the predicted cooling rates diverge at other times.

In summary, the Cole model differs from the previously
introduced models in three ways.
First, it resets the halo birth time at discrete points
according to the mass accretion history.
Second, it redistributes the gas only at these reset points, instead of continuously.
Third, it adopts a gas density profile with a large core.
To make fair comparisons to
the other models and to understand the effect of the first two modifications, we also run
models using a pure singular isothermal profile (eq. \ref{equ:isoden})
for the gas. In this case, the cooling rate is
\begin{equation}\label{equ:cr_cole_iso}
\mcooldot={m_{\rm gas}(r_{\rm min}) -m_{\rm gas}'(r'_{\rm min})
\over \Delta t} = 4\pi\rho_0 \rvir^2 { r_{\rm min} - r'_{\rm min} \over \Delta t},
\end{equation}
where
\begin{equation}
\rho_0={m_{\rm gas0} \over 4\pi r_{\rm vir}^3}.
\end{equation}

\subsection{The Somerville Model}\label{subsec:some}

The Somerville model is similar to the Cole model, but instead of relying on
a factor of two increase in halo mass to redistribute the hot gas, it redistributes
the gas whenever a major merger occurs with the primary halo.  Therefore, a significant mass
enhancement takes place in a time step. Following \citet{Somerville1999a}, we
reset the birth time when a halo accretes an amount of mass that is larger
than its own in a single step. When a merger with a smaller mass ratio
(a minor merger) occurs, any hot gas associated with the secondary halo is 
added to the primary halo outside of the cooling radius. 
The shape of the halo gas density profile does not change, 
but the normalisation is increased. The halo lifetime, $t_{\rm age}$, 
is either the time since the top-level of the merger tree began or
the time since the last major merger, whichever is shorter.  
The cooling radius is then defined by
$\tau_{\rm cool}=t_{\rm age}$.  In addition, any centrally accreted gas 
must be within 
a sound-speed radius, $r_{\rm ss} = c_{\rm s} t_{\rm age}$.
The sound speed
$c_{\rm s}=(5kT/3\mu m_{\rm p})^{1/2} \sim 1.3\sigma_{v} \approx 0.9 V_c$,
where $\sigma_{v}$ is the 1D velocity dispersion of the halo.
Effectively, any hot gas enclosed by a radius
$r_{\rm min} = \min[r_{\rm cool}, r_{\rm ss}, \rvir]$,
can cool. Thus, the cooling rate is defined as in equation
(\ref{equ:cr_cole_iso}) with a modified normalisation.  The new
gas-mass normalisation,
\begin{equation}
\rho_0={m_{\rm gas} \over 4 \pi \rvir^2 (\rvir - r'_{\rm min})},
\end{equation}
accounts for the new mass deposited by minor mergers between
the cooling radius and the virial radius.
If the cooling radius is smaller than the sound-speed radius and the virial radius,
then cooling is in the ``slow'' regime; otherwise cooling is in the ``rapid'' regime.

\subsection{The MORGANA Model}\label{subsec:morg}
The MORGANA model was first described in \citet{Monaco2007}, and 
its cooling model was studied in more detail in \citet{Viola2008}. 
Readers are referred to those papers for details of the model. 
The MORGANA model differs from the other models mainly in two aspects. 
First, it does not
assume that hot gas transfers to a cold phase after a cooling 
time. Second, although it also assumes a cooling radius, which separates 
the inner cooled gas from the outer hot halo gas, the treatment of the 
cooling radius is different from in the other models. 
In the MORGANA model, all the gas shells contribute to cooling mass 
and the cooling rate is calculated as an integral over radius. 
The model starts with a gas density profile and temperature profile. 
Using equation\ \ref{equ:cooltime}
one can then calculate the cooling time, $t_{\rm cool}(r)$, 
for every gas shell. 
The cooling rate of a gas shell at radius $r$ is 
\begin{equation}\label{equ:coolrate_morg}
\Delta \mcooldot(r) = {4 \pi r^2 \rho_{\rm gas}(r) \over t_{\rm cool}(r)} \Delta r. 
\end{equation}
The MORGANA model also assumes a sharp border in radius separating the cooled 
gas from the hot halo gas, so that
the total cooling rate is calculated by integrating 
the contribution over radii ranging from the cooling radius, $r_{\rm M}$,  
to the virial radius. The cooling radius marches 
outwards as more gas cools from the hot halo. The time derivative of the 
the cooling radius is 
\begin{equation}\label{equ:coolradius_morg}
\dot{r}_{\rm M} = {\mcooldot \over 4 \pi \rho_{\rm gas}(r_{\rm M}) r_{\rm M}^2}. 
\end{equation}
Moreover, also differing from other models, the MORGANA model assumes 
a gas density profile and temperature 
profile that are in hydrostatic equilibrium in an NFW halo. To solve for 
the cooling rate in an NFW halo requires one to compute the integral of 
equation\ \ref{equ:coolrate_morg} numerically. The appendix 
of \citet{Viola2008} derives an analytic solution for the cooling rate 
of a model with a singular isothermal density profile for the gas 
when $r_{\rm M} << \rvir$. 
In such a case, the cooling rate at time, $t$, since the halo formed is 
\begin{equation}\label{equ:coolrate_miso}
\mcooldot(t) = { m_{\rm gas} \over t_{\rm c,v}} \sqrt{t_{\rm c,v} \over 2 t},
\end{equation}
where $t_{\rm c,v}$ is the cooling time at the virial radius of the halo, 
and the cooling radius is 
\begin{equation}\label{equ:coolradius_miso}
r_{\rm M}(t)= \rvir \sqrt{ 2 t \over t_{\rm c,v}}.
\end{equation}
The formation time of a halo is defined as the time when the halo had 
the last major merger.
In addition, the MORGANA model assumes that the increase in the cooling radius 
is counteracted by a shrinking at the sound speed, 
e.g. 
\begin{equation}\label{equ:coolradius_ch}
\dot{r'}_{\rm M}(t) = \dot{r}_{\rm M}(t) - c_{\rm s}(r'_{\rm M,ch}),
\end{equation}
where $r'_{\rm M}$ is the corrected inner radius of the hot gas 
and $c_{\rm s}$ is the sound speed. 
The model assumes that the cooled gas accretes to the central galaxy at a rate
\begin{equation}
\dot{m}_{acc} = {m_{\rm cool} \over n_{\rm dyn} \tau_{\rm dyn}(r'_{\rm M})},
\end{equation}
where $\tau_{\rm dyn}(r'_{\rm M})$ is the dynamical time of the halo 
at $r'_{\rm M}$ and $n_{\rm dyn}$ is a free parameter that controls the delay 
in the accretion. 
To make our comparison study across all the models on a fair basis,
instead of implementing the original MORGANA model, 
we adopt the solution for the isothermal density profile. Furthermore, 
we also neglect the shrinking effect of $r_{\rm M}$ and the delay of accretion 
in our implementation 
by simply setting $r'_{\rm M}=r_{\rm M}$ and $\dot{m}_{acc} = \dot{m}_{\rm cool}$. 
Note that the original model does not differentiate the ``rapid''
and ``slow'' accretion modes. 

\section{Model Comparison}\label{sec:res}

\subsection{Cooling in Smoothly Accreting Haloes}\label{subsec:shalo}

To make predictions for radiative cooling in a dark matter halo, one
needs to incorporate a prescription of gas cooling into the halo
formation process. In conventional semi-analytic models, the formation
of a halo is described by its merging tree. However, before we 
examine the performance of the cooling algorithms using more
realistic case using halo merger trees, we first examine a simple model
to gain some insight. In this model, dark matter haloes are assumed to
grow purely through smooth, spherical accretion. We model the mass
accretion histories (MAHs, hereafter) of dark matter haloes
using the formula of \citet{Wechsler2002} obtained by fitting results
of $N$-body simulations:
\begin{equation}\label{equ:mah}
M(a)=M_0 \exp\left[-2 a_c \left({a_0\over a}-1\right)\right],
\end{equation}
where $a$ is the expansion scale factor, $a_c$ is the scale factor
corresponding to the formation time of the halo, and $M_0$ is the mass of the
halo at the time of observation $t_0$ (corresponding to $a_0$). The factor
$a_c$ is the only free parameter that characterises the
shape of a MAH. We investigate four cases, with halo masses at the
present time ($z=0$) of $\mvir=10^{11}, 10^{12}, 10^{13}$ and $10^{14}\msun$, 
respectively. These masses roughly cover the galaxy formation mass
range. In a CDM universe, haloes with
smaller masses on average form earlier, and so we use a mass-dependent
$a_c$ obtained from $N$-body simulations
\citep{Bullock2001a, Wechsler2002, Zhao2003a}: 
$a_c=0.45, 0.53, 0.6$ and $0.75$
for haloes with $\mvir=10^{11}, 10^{12}, 10^{13}$ and $10^{14}\msun$,
respectively. With these values, the resulting MAHs closely match the average
MAHs obtained from the Monte-Carlo merger trees based on  the
extended Press-Schechter (EPS) formalism.  Owing to the 
finite numerical resolution of the simulation, the
cooling histories are not reliable before $z\approx 4$. 

%dln(1+z) = 0.07 at z=0, <0.01 at high z
We follow the mass growth of a halo using this smoothed MAH
from $z=7$ to the present time using 100 steps equally spaced in $z$. The time step is small
enough to resolve all the processes important to our study.
We include radiative cooling using all the cooling models
presented in  \S\ref{sec:model} except the Somerville model, which
requires individual merger events to assign halo lifetimes.
For the other models,
we follow the halo growth for each of the four halo masses,
and calculate the cooling rate and the total cold
gas mass at each time step.
For the MORGANA model, we adopt the time since the beginning 
of the simulation as the age of the halo. We use this halo age definition to compute the smooth mass accretion history in this section.
For comparison, we also use a spherically symmetric, hydrodynamic simulation code
\citep{Lu2007} to follow the cooling histories.  
The initial conditions are chosen so that haloes
exactly reproduce the mass accretion histories given by
equation (\ref{equ:mah}).
Readers who are interested in the details
of the simulation setup are referred to \citet{Lu2006} and \citet{Lu2007}.
We use $10^5$ equal-mass shells for the dark matter and 5,000
equal-mass shells for the gas.  Only half of the
mass lies within the virial radius at the present time to minimise the effects of the
outer boundary.
We choose
the same cosmological parameters and use the same cooling rates
as for the SAMs. 

\begin{figure}
  \hfill
  \begin{minipage}[t]{.45\textwidth}
    \begin{center} 
      \epsfig{file=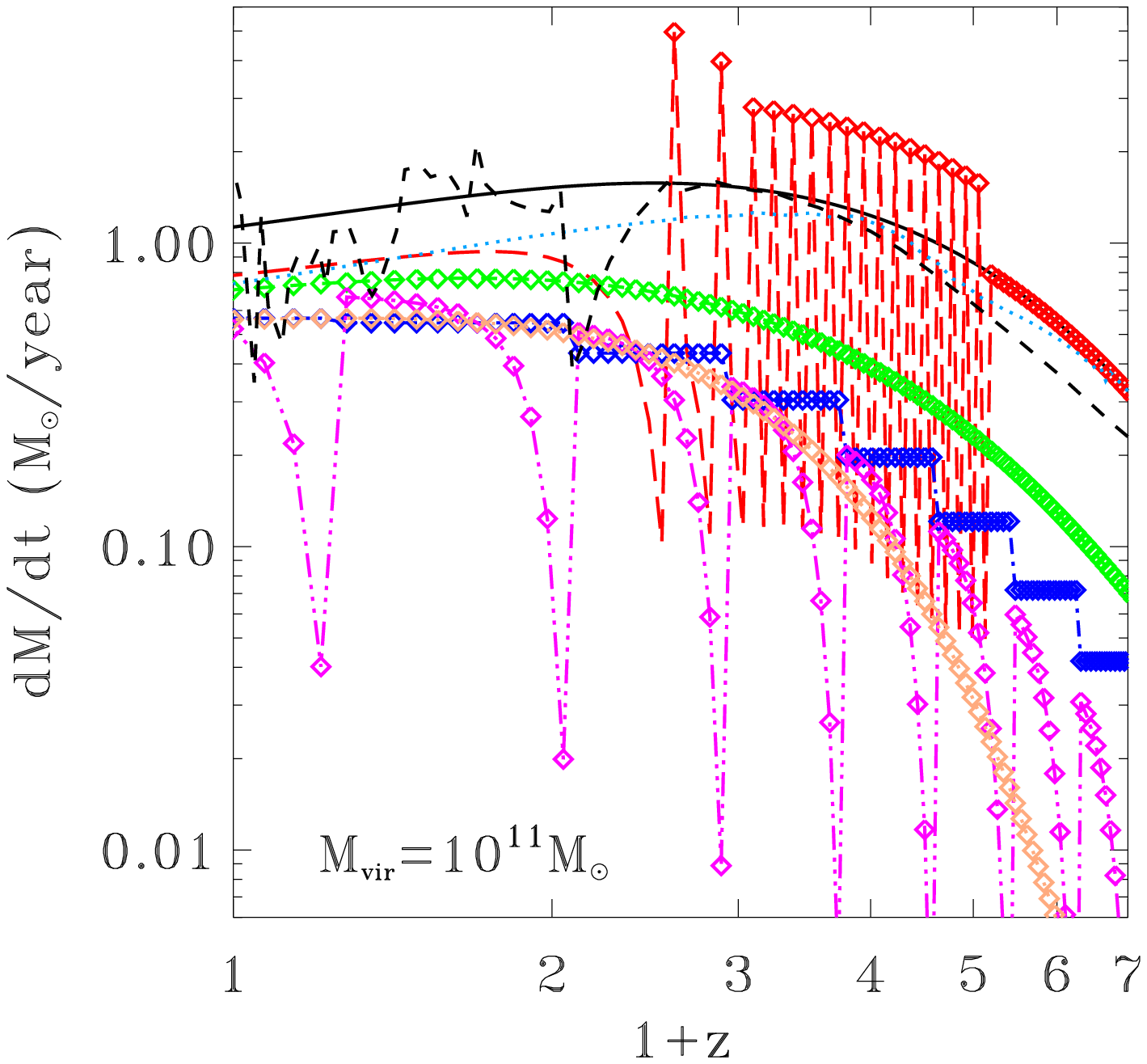, scale=0.5}
    \end{center}
  \end{minipage}
  \hfill
  \begin{minipage}[t]{.45\textwidth}
    \begin{center} 
      \epsfig{file=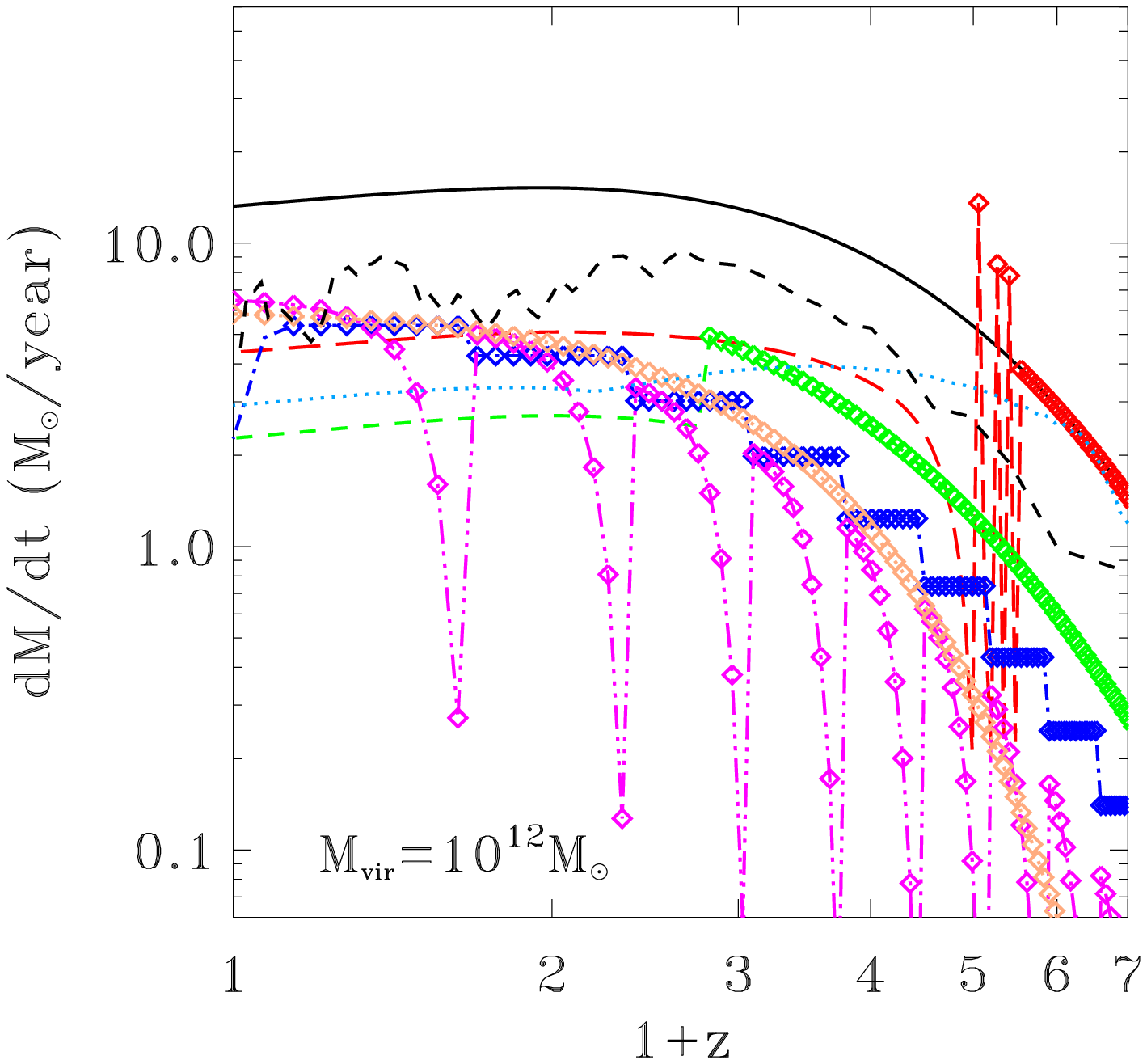, scale=0.5}
    \end{center}
  \end{minipage}
  \hfill

 \vfill
  \hfill
  \begin{minipage}[t]{.45\textwidth}
    \begin{center}
      \epsfig{file=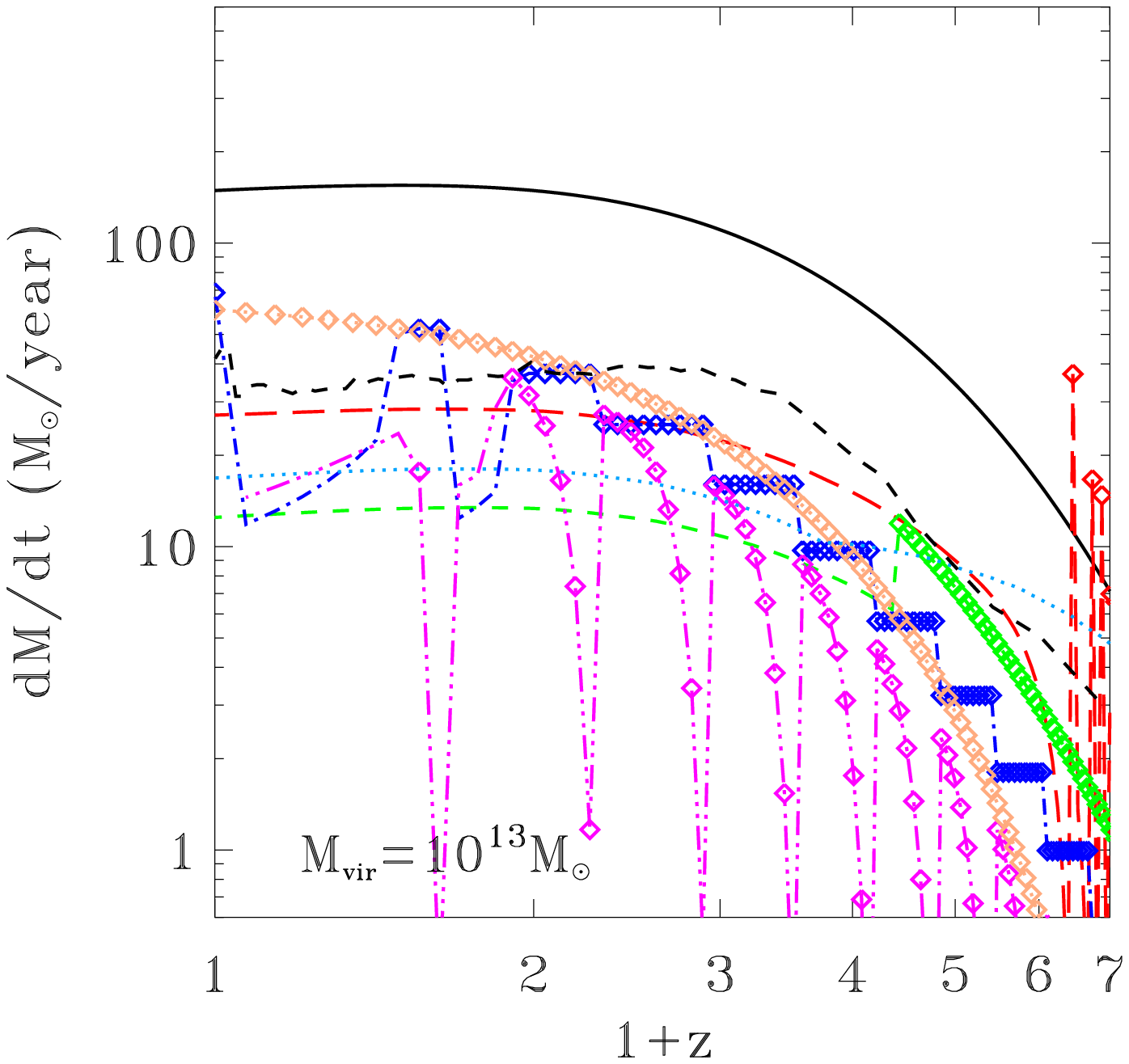, scale=0.5}
    \end{center}
  \end{minipage}
  \hfill
  \begin{minipage}[t]{.45\textwidth}
    \begin{center}
      \epsfig{file=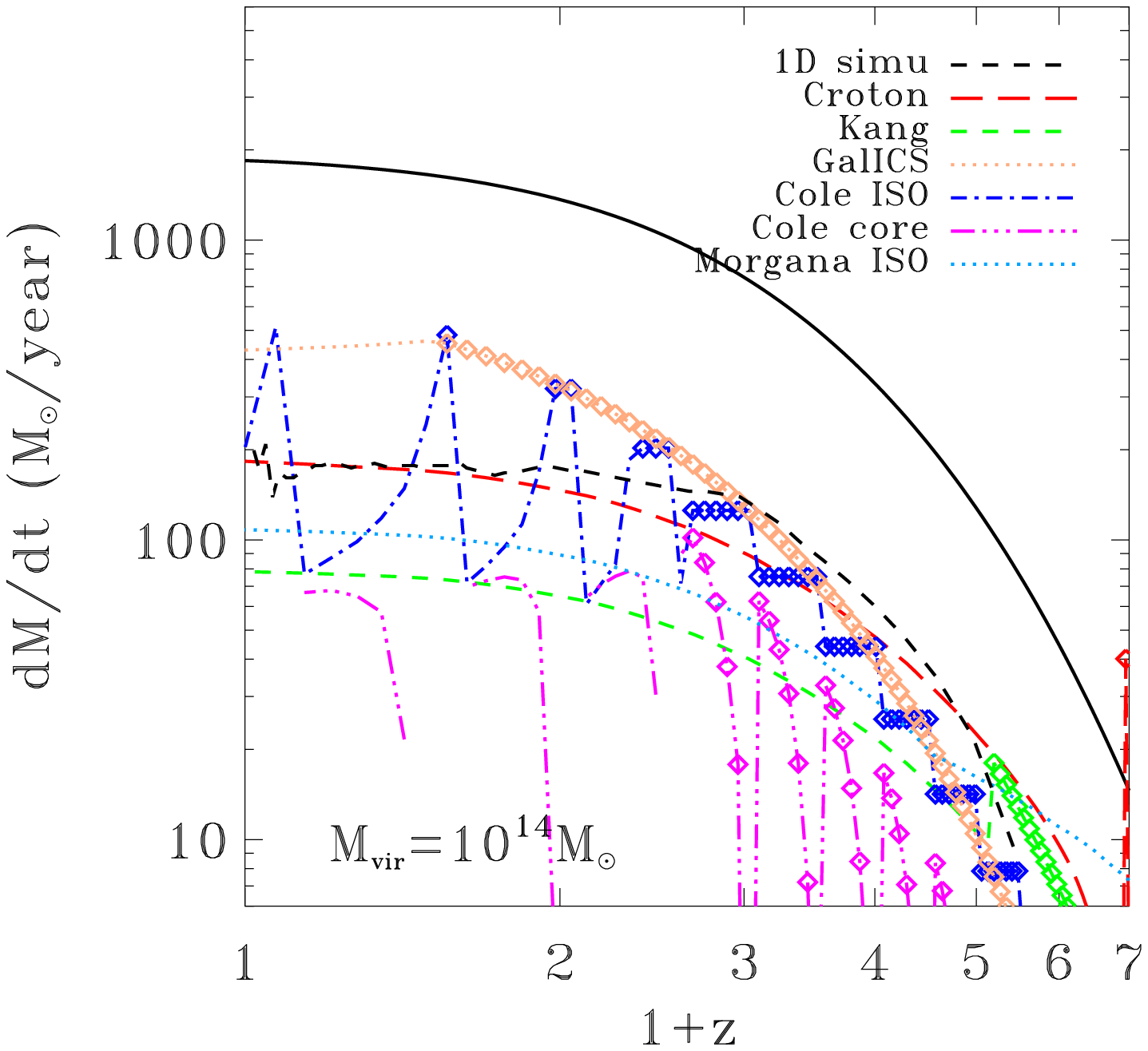, scale=0.5}
    \end{center}
  \end{minipage}
  \hfill
\caption{\large The cooling rates of dark matter haloes
with a smooth, averaged accretion history.
The panels show results with
different final virial masses ($10^{11}, 10^{12}, 10^{13}$ and $10^{14}\msun$)
as labeled.
The solid line denotes the rate
at which baryons accrete into the dark matter halo. Each of the other lines
denotes the
cooling rate predicted by a cooling algorithm, as noted in the upper-right
corner of the forth panel (see text). The segments of the lines covered by diamonds
represent the ``rapid'' cooling regime and the plain lines
represent the ``slow'' cooling regime. (See \S\ref{sec:model} for
the cooling regime identification in each model.) 
Note that the MORGANA model does not differentiate the cooling 
regimes.
}\label{fig:scr}
\end{figure}

\begin{figure}
  \hfill
  \begin{minipage}[t]{.45\textwidth}
    \begin{center}
      \epsfig{file=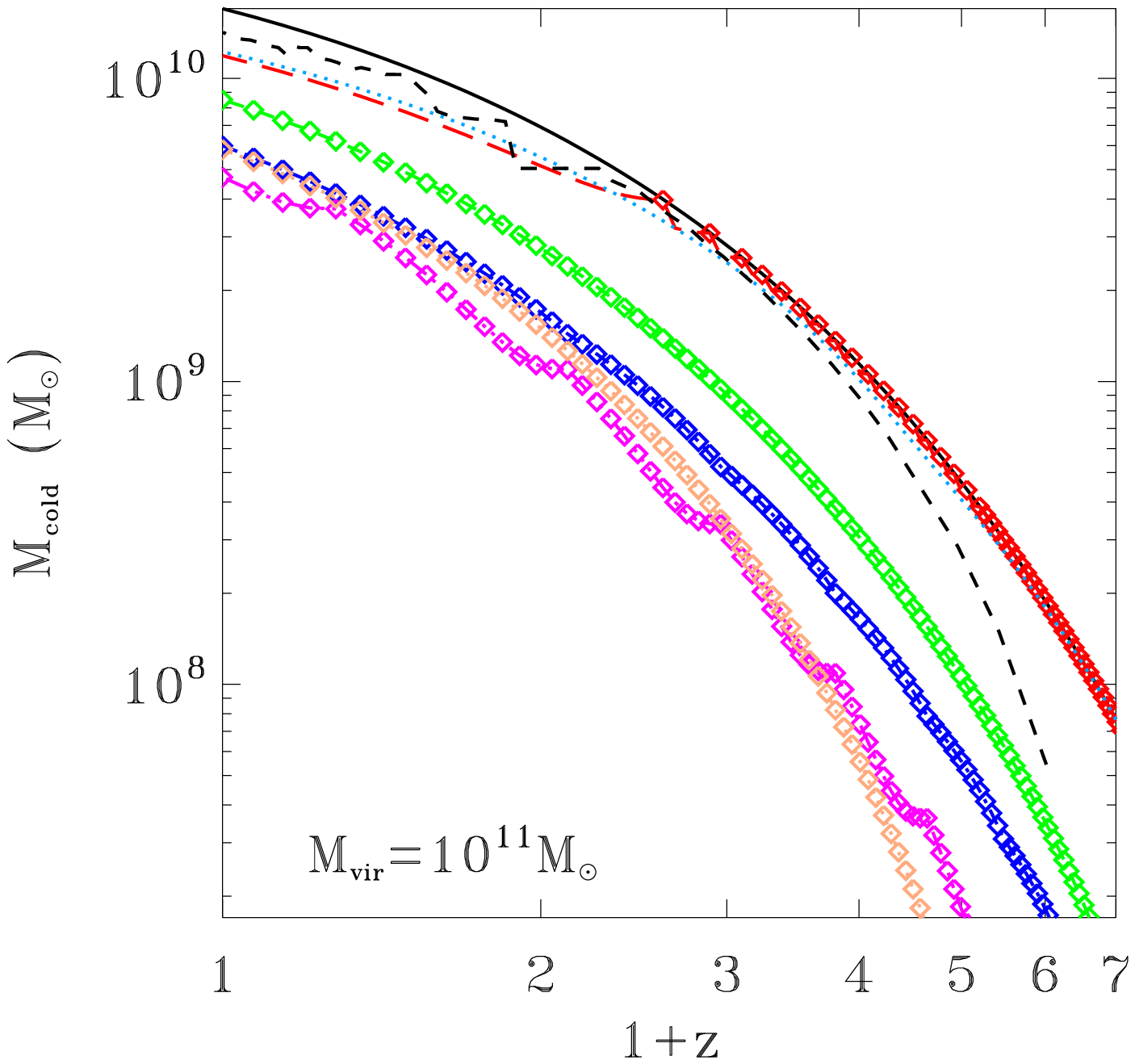, scale=0.5}
    \end{center}
  \end{minipage}
  \hfill
  \begin{minipage}[t]{.45\textwidth}
    \begin{center}
      \epsfig{file=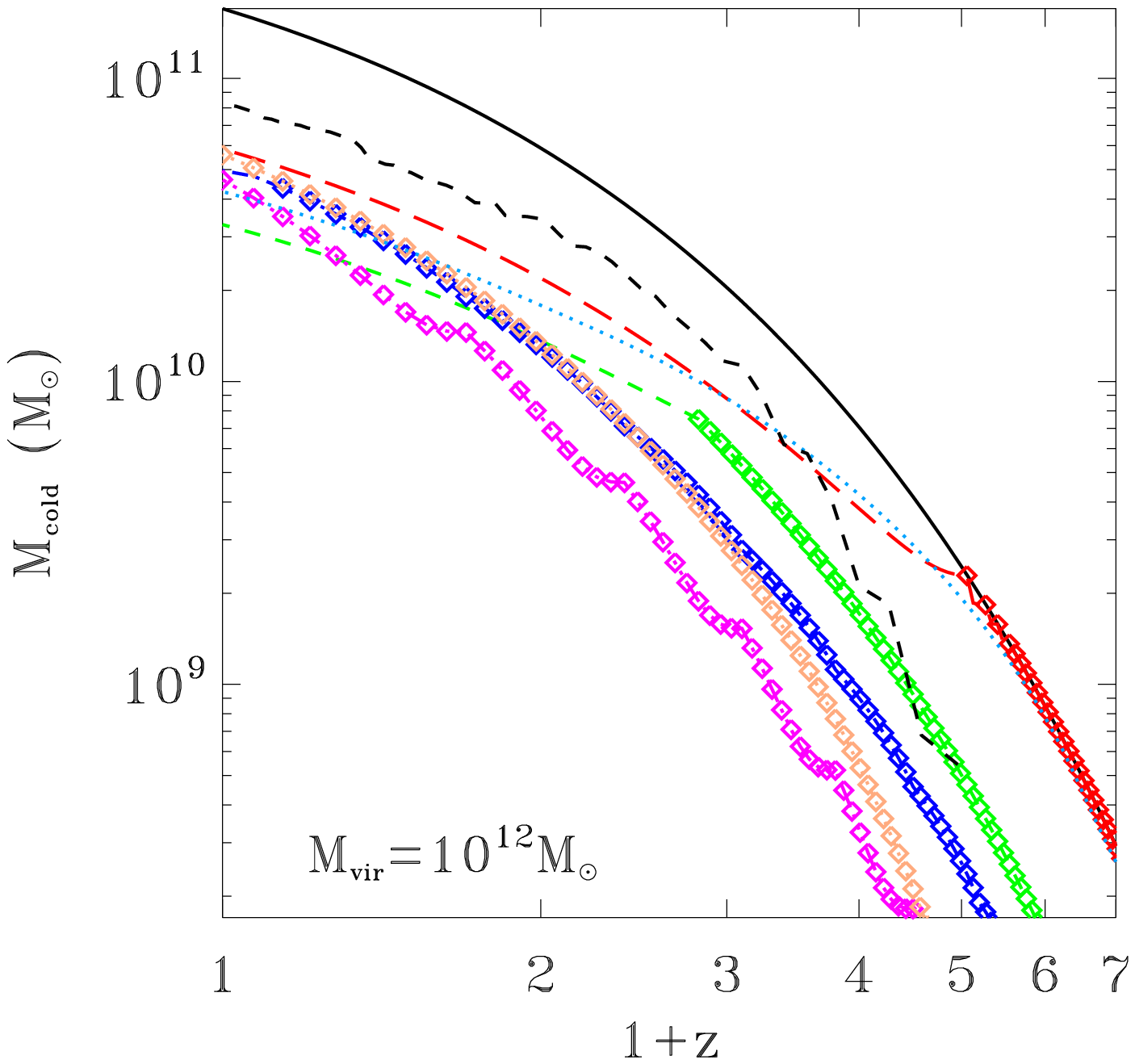, scale=0.5}
    \end{center}
  \end{minipage}
  \hfill

 \vfill
  \hfill
  \begin{minipage}[t]{.45\textwidth}
    \begin{center}
      \epsfig{file=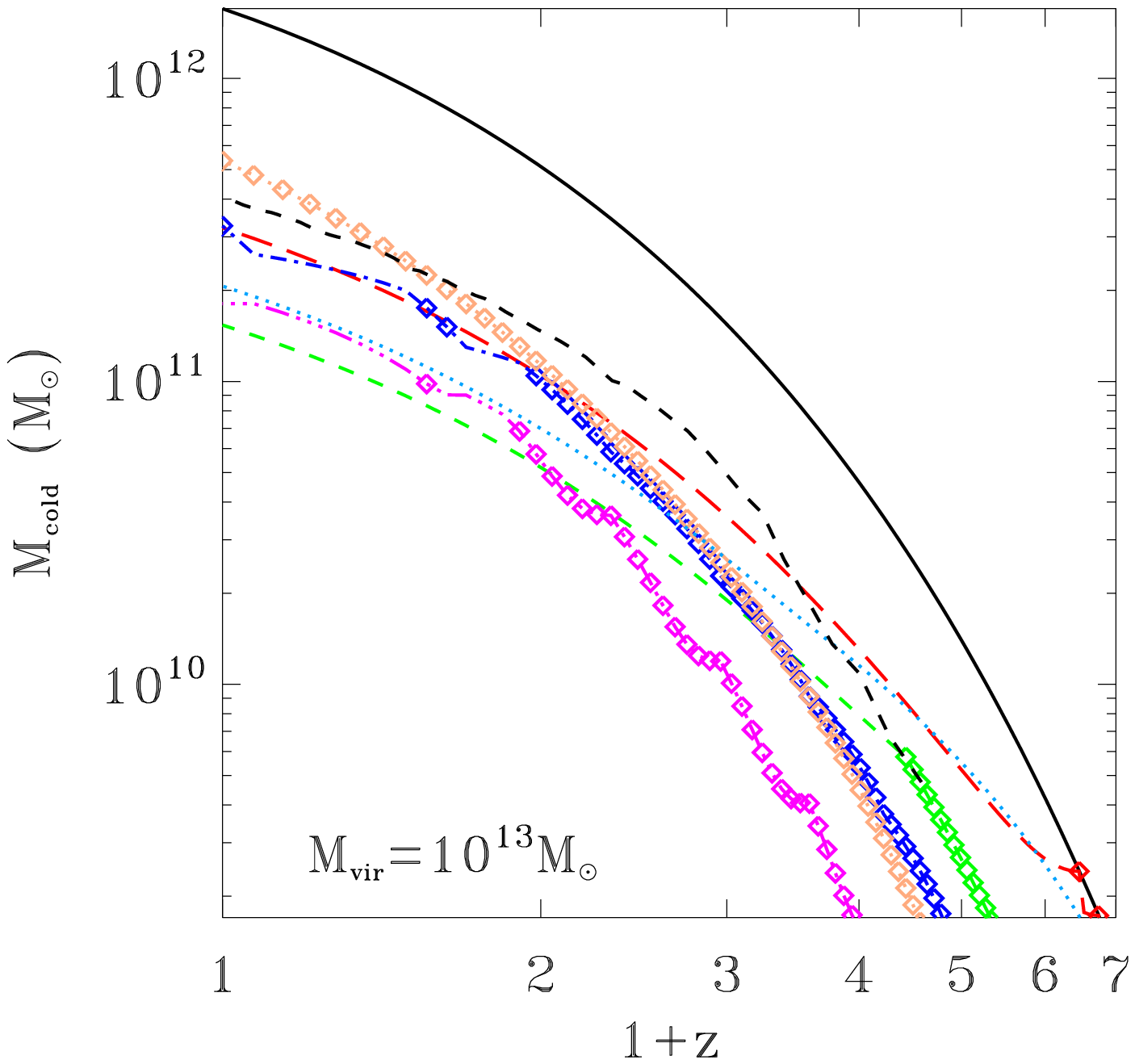, scale=0.5}
    \end{center}
  \end{minipage}
  \hfill
  \begin{minipage}[t]{.45\textwidth}
    \begin{center}
      \epsfig{file=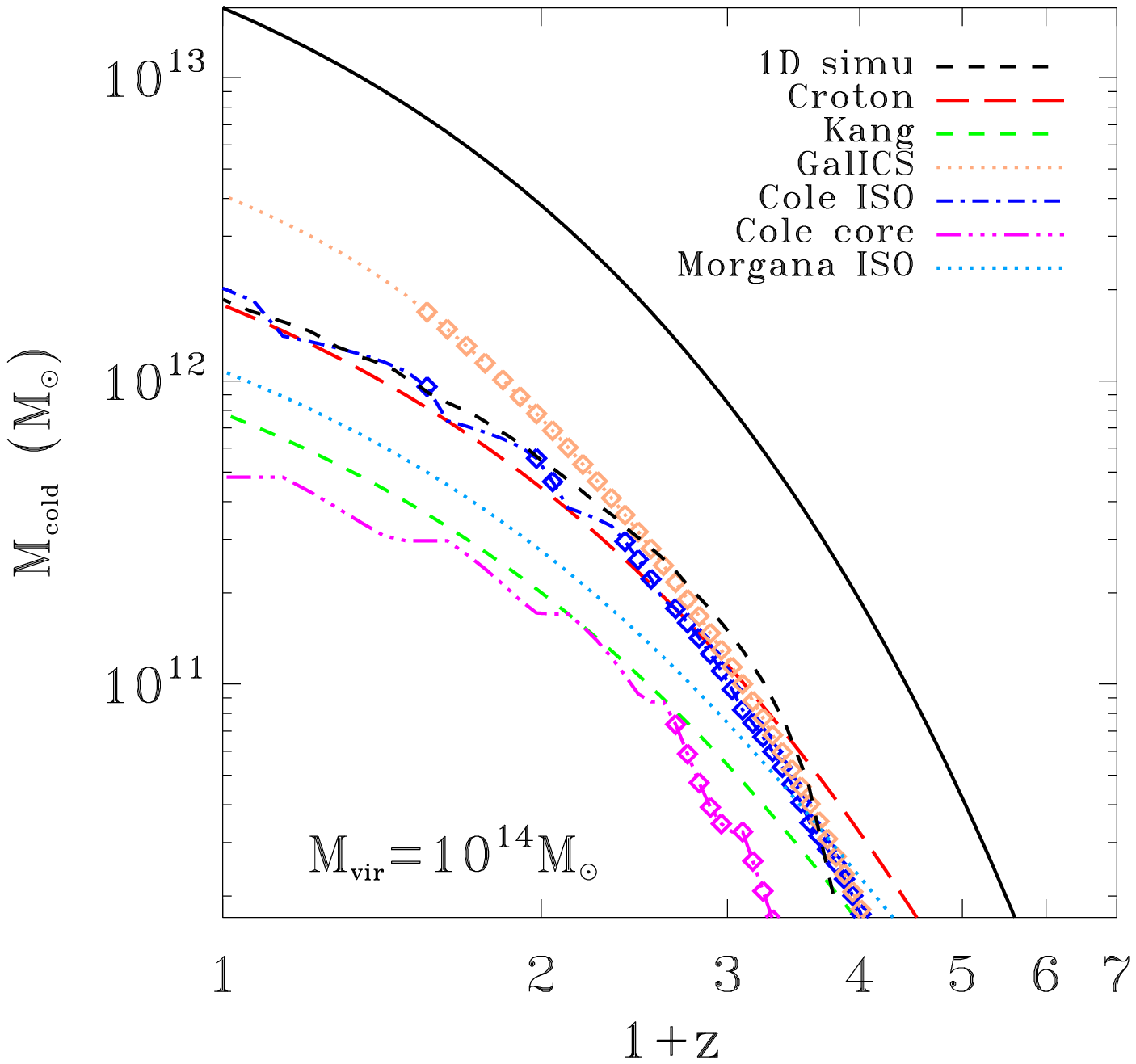, scale=0.5}
    \end{center}
  \end{minipage}
  \hfill
\caption{\large The accumulated cold gas mass in the central
galaxy for the same model as Fig.~\ref{fig:scr}, using the
same halo masses.
The solid lines plot the total mass of baryons in the dark matter halo.
The black dashed lines plot the spherically symmetric simulation results.
}\label{fig:scm}
\end{figure}

Figures~\ref{fig:scr} and \ref{fig:scm} show the predicted 
cooling rates and the accumulated cold gas masses, respectively.
In these plots, we denote cooling in the ``rapid'' regime by
diamonds and that in the ``slow'' regime by lines. The black short-dashed line
in each panel shows the result of the spherical simulation.  As one
can see, in all the models, low-mass haloes and high-redshift haloes are dominated by ``rapid''
cooling while massive haloes at low redshift tend to be dominated by the ``slow'' cooling
phase. However, since different models have different criteria for the cooling modes,
the predicted transition times between these two cooling regimes are
very different in the different models,  and the halo
masses at the transition can vary by almost two orders of magnitude.
Thus, the transition mass in SAMs in general  does not correspond to
a given halo mass scale where shock heating of the gas becomes important
\citep[see also discussion in Appendix of][]{Kerevs2005}.

The different models predict different cooling histories
for the same dark matter halo. Since the Croton, Kang, and GalICS 
models redistribute the gas at every time step, and since the gas supply 
in the smooth-accretion halo model is continuous, these three 
models predict smooth cooling histories in both the ``rapid'' and ``slow'' 
cooling regimes. Since the Croton model accretes all the halo gas onto 
the central galaxy  when the halo is in the ``rapid'' mode, the cooling 
rate follows the halo accretion rate in the early history of the halo. 
When haloes are in transition to the ``slow'' cooling regime, however, 
the accretion rate fluctuates as the haloes switch modes in adjacent 
time steps (see \S\ref{sec:model}). In the Kang model, there is a factor 
of two drop in the cooling rate at this transition,
as expected from their 
parameterisations (also see \S\ref{sec:model}). The Croton model predicts 
a higher cooling rate than the Kang model, because 
the former assumes a shorter timescale for gas cooling.
However, the reduced cooling 
in the Kang model leaves more gas in the halo, which enhances the cooling 
rate ($\propto m_{gas}^2$), thereby reducing the difference between the two models. 
This explains why the predicted total amounts
of cooled gas for low-mass haloes in these models become more similar at late times.

Since the gas distribution is reset when the halo
mass is doubled in the Cole model, the gas supply is not continuous. Thus, the cooling rate shows
discrete features after every halo birth time. If one replaces the gas
distribution with a singular isothermal density profile in the Cole
model, the
gas suddenly forms a central density cusp every time the halo mass
doubles, boosting the cooling rate.
When a halo is in the ``rapid'' cooling regime, the free-fall time determines the
cooling rate. Hence, in the singular isothermal profile case, the cooling rate
remains constant until the halo doubles its mass and the free-fall
time changes. However, when the halo is in the ``slow'' 
cooling regime, the cooling radius determines the cooling rate, 
and the cooling rate drops with time owing to the decrease in 
the local gas density. When a core is added to the halo gas density 
distribution and the halo gas is redistributed when 
the halo mass doubles,
the cooling rate then drops because 
the central gas density is low and the time interval $t-t_{\rm form}$
available for the hot halo gas 
to cool in the newly formed halo is short. However, as the cooling radius marches 
outwards, the local density approaches that of the singular isothermal 
density profile and the cooling rate approaches
that seen in the model without a core. For very massive haloes with a cored 
gas density profile, the cooling rate can be very low owing to their 
high temperatures and low central densities \citep[also see][]{Viola2008}. 

If the GalICS model adopted an singular isothermal gas profile, its cooling
rate would follow the upper envelope of
the core-free Cole model prediction. Since the model has a
small core radius, the predicted  cooling rate is right between the
Cole models with and without a core. This is not surprising given that 
the GalICS model adopts the same formula to predict the cooling rate 
as the Cole model right after a halo doubles its mass, as discussed 
in \S\ref{subsec:cole}. 

Since the MORGANA model does not distinguish the ``rapid'' and the ``slow'' 
cooling regimes, in Figure \ref{fig:scr} and \ref{fig:scm} we only plot 
the predicted total cooling rates and cold gas masses (the light blue dotted lines). 
The model, with the assumption of an isothermal gas profile, 
generally predicts higher cooling rates at early times for all halo masses. 
This characteristic has also been pointed out by previous studies using static halos 
and realistic gas profiles 
\citep{Viola2008}. At late times, the predicted cooling rates of 
the MORGANA model for an isothermal profile are 
about 1.4 times higher than those of the Kang model, 
which is the closest to the classic \citet{White1991} model among the models. 
This is in an excellent agreement with the analytic solutions of 
the two models for the isothermal gas profile presented in the appendix of \citet{Viola2008}, 
\citep[also see][for a more detailed discussion]{Bertschinger1989, Keres2007}. 

To compare with the Somerville model,
one might consider assigning the halo lifetime as the Hubble time, rather
than the time to the last merger.  
Without major mergers and because halo gas cools efficiently at
high redshift, the model would predict that all the accreted gas would cool in a halo
regardless of its mass. 
Without a new halo ``birth time'' to redistribute the gas, the model
would always add the newly accreted gas to a thin shell between the cooling radius
and the virial radius.  Therefore, the halo gas density would remain high and
cool efficiently throughout its lifetime. 

As seen in Figure \ref{fig:scm},  for the two lower-mass haloes, the total cold 
gas masses predicted by all the models are within a factor of about 2 of each 
other at the present time. The differences increase with mass 
and are the largest for the $10^{14}\msun$ halo.
For these massive haloes, the GalICS model predicts the largest amount
of cold gas; the Cole model without a core and the Croton model are in the middle;
and the Cole model with a core and the Kang model predict the smallest amounts.
Although the cooling rates predicted by the two Cole models look very different from
the other models owing to their discontinuous behaviour, the total cold gas
masses predicted by the Cole model without a core for haloes with masses 
$10^{12}$--$10^{14} \msun$ are quite close to the predictions of the Croton model.
For the two intermediate mass haloes, the total cold gas mass predicted by the GalICS model
closely tracks that of the Cole model without a core over all
redshifts, but for the higher mass haloes the cold gas mass predicted
by GalICS model is larger than that of the Cole model at redshifts
$z\la 2$. Compared with the spherically symmetric, hydrodynamic simulation, all the
models under-predict the cooling rates for small haloes at late times,
with only the Croton model approaching the simulation result.
For massive haloes, the Croton and Cole models without a core are consistent with the
simulation in the ``slow'' cooling regime.

%%%%%%%%%%%%%%%%%%%%%%%%%%%%%%%%%%%%%%%%%%%%%%%%%%%%%%%%%%%%%%%%%%%

\subsection{Cooling in Merging Haloes}\label{subsec:mhalo}

Cold dark matter haloes form hierarchically through both accretion and
the merging of smaller structures. Therefore, the models presented in the
previous section, which are based on the assumption of  smooth
accretion, do not fully capture the realistic picture of gas cooling in CDM haloes.
In this section, we incorporate the different cooling algorithms into
merger trees to show the performance of these algorithms under more
realistic conditions.

Halo merger trees can either be extracted from $N$-body simulations or
generated using Monte-Carlo methods. Because merger trees from N-body 
simulations contain information about both halo dynamics and environment, 
they have been widely adopted lately for modelling galaxy formation. 
However, Monte-Carlo merger trees,  which lack this information,  
are still a powerful alternative as they are much easier to generate 
and have infinite resolution in principle. In this section, we adopt 
Monte-Carlo merger trees generated with the algorithm recently developed 
by \citet{Parkinson2008}. This algorithm was tuned to match the conditional 
mass functions from $N$-body simulations.  For a simple, illustrative example, 
we choose the free parameters in the model so that the resulting halo 
conditional mass function matches the EPS conditional mass function, i.e. 
$G_0=1$ and $\gamma_1=\gamma_2=0$ \citep[see][]{Parkinson2008}.
The virial radius of a halo is defined so that the mean over-density within 
it is a factor $\delta_{\rm vir}$ times the critical density of the universe.
We use the fitting formula by \citet{Bryan1998} to calculate
$\delta_{\rm vir}$. This definition of virial radius is also the same
as that used in the simulations presented in this paper.

As in the previous section, we study four cases, with halo masses of 
$\mvir=10^{11}, 10^{12}, 10^{13}$ and $10^{14}\msun$ at the present time.
It is worth noting that the main difference between the merger tree model
and the smooth accretion model is that gas can cool in all progenitor haloes
in a merger tree but only cool in the main branch haloes in the smooth accretion model.
To make a fair comparison with the results obtained
in the last subsection, we only look at the main branch of the
halo merger trees. The cooling histories
of different haloes can be very different even if they have the same
final mass
because for different haloes mergers occur at various times
and with different progenitor mass ratios.
To sample the cooling histories, we generate 100
merger trees for each final halo mass and average the results.
We set the mass resolution (the minimum halo mass tracked
by the merger tree) to be 0.001 times the final mass of the halo. 
We store merger trees at 60 redshifts that are equally spaced   
in $\log(1+z)$ from $z=6$ to $z=0$.
We will test the effects of the choice of mass 
and time resolution on the predicted cooling histories, 
and demonstrate that our results are numerically robust to these choices. 

\begin{figure}
  \hfill
  \begin{minipage}[t]{.45\textwidth}
    \begin{center}
      \epsfig{file=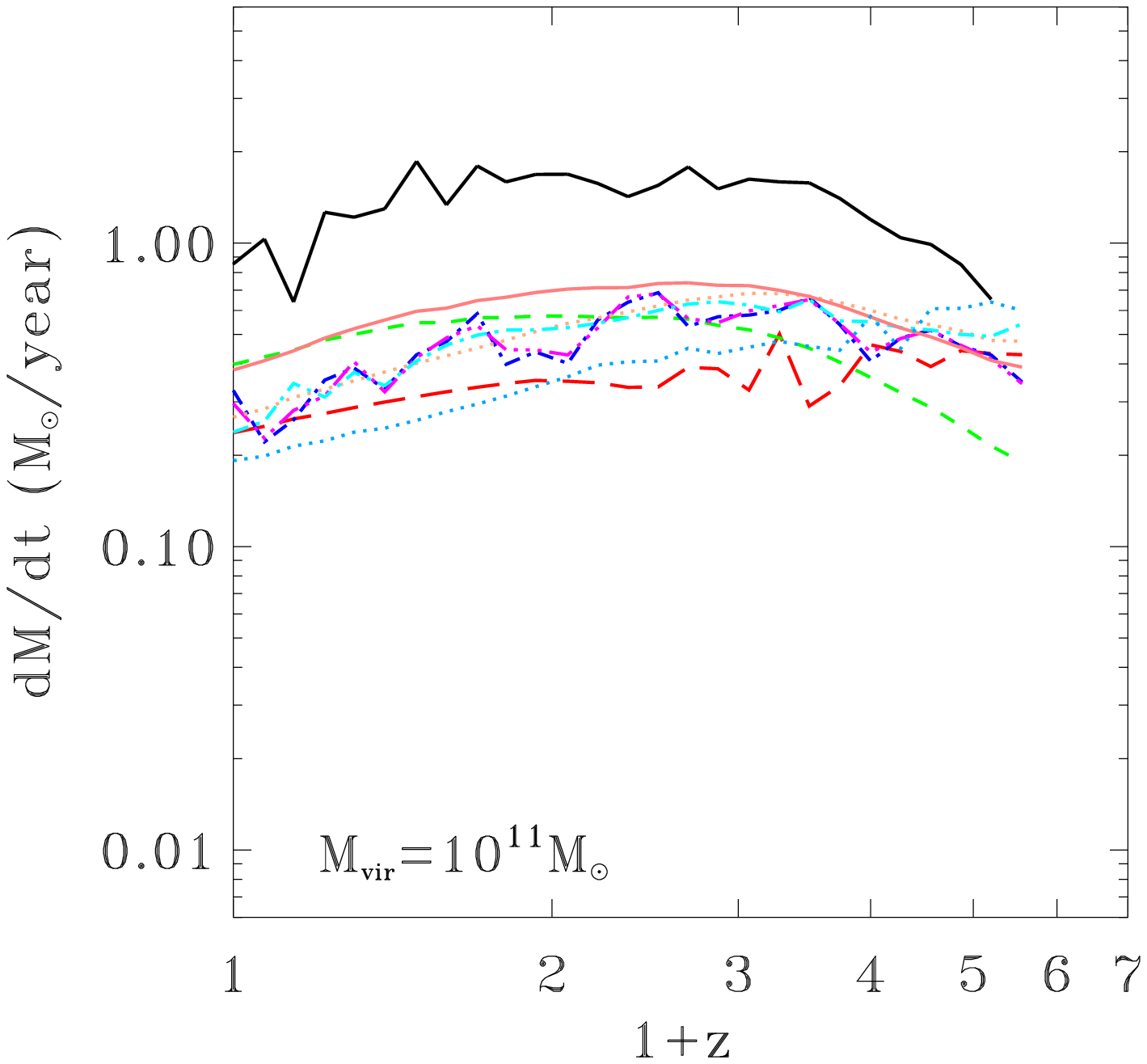, scale=0.5}
    \end{center}
  \end{minipage}
  \hfill
  \begin{minipage}[t]{.45\textwidth}
    \begin{center}
      \epsfig{file=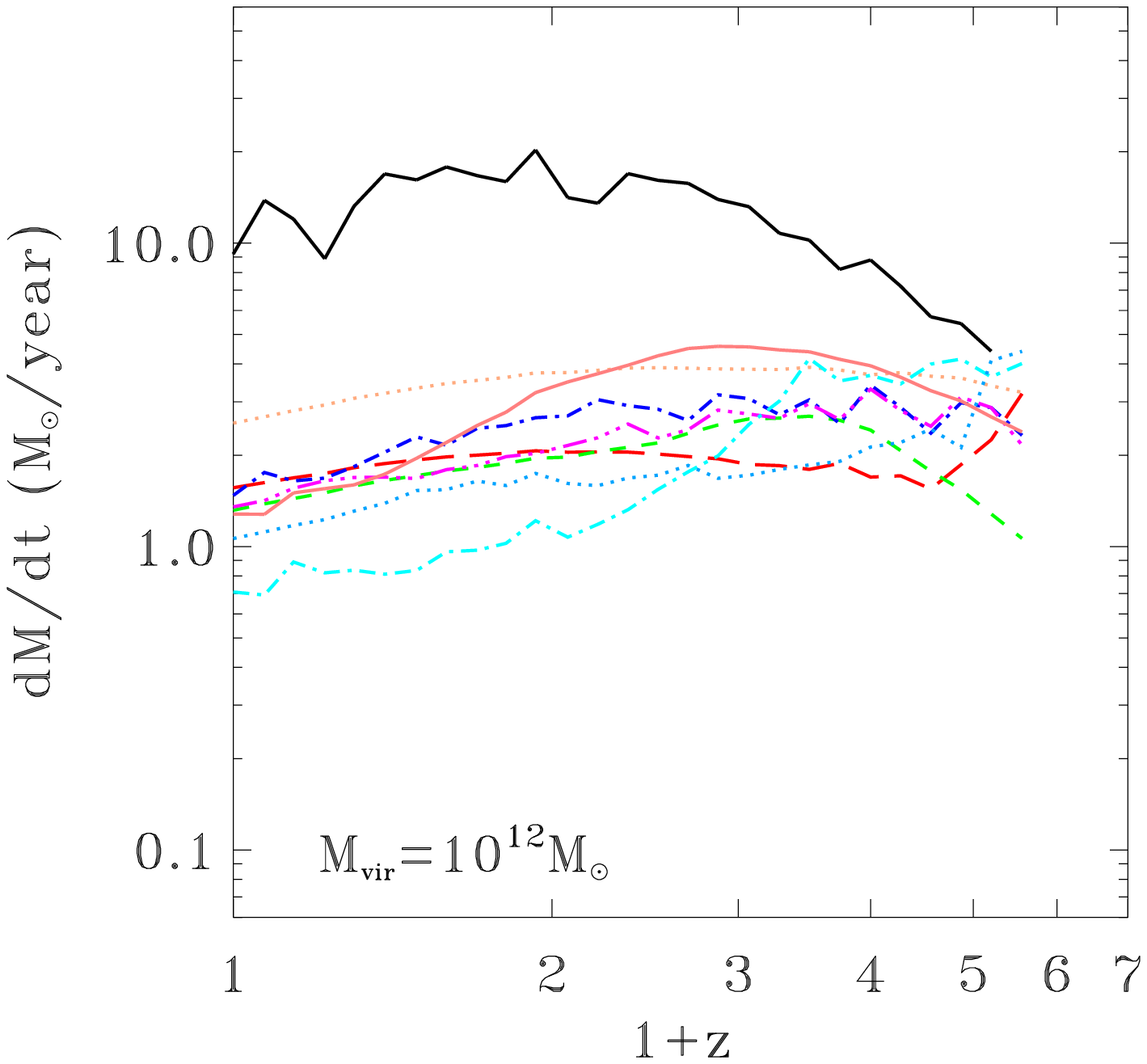, scale=0.5}
    \end{center}
  \end{minipage}
  \hfill

 \vfill
  \hfill
  \begin{minipage}[t]{.45\textwidth}
    \begin{center}
      \epsfig{file=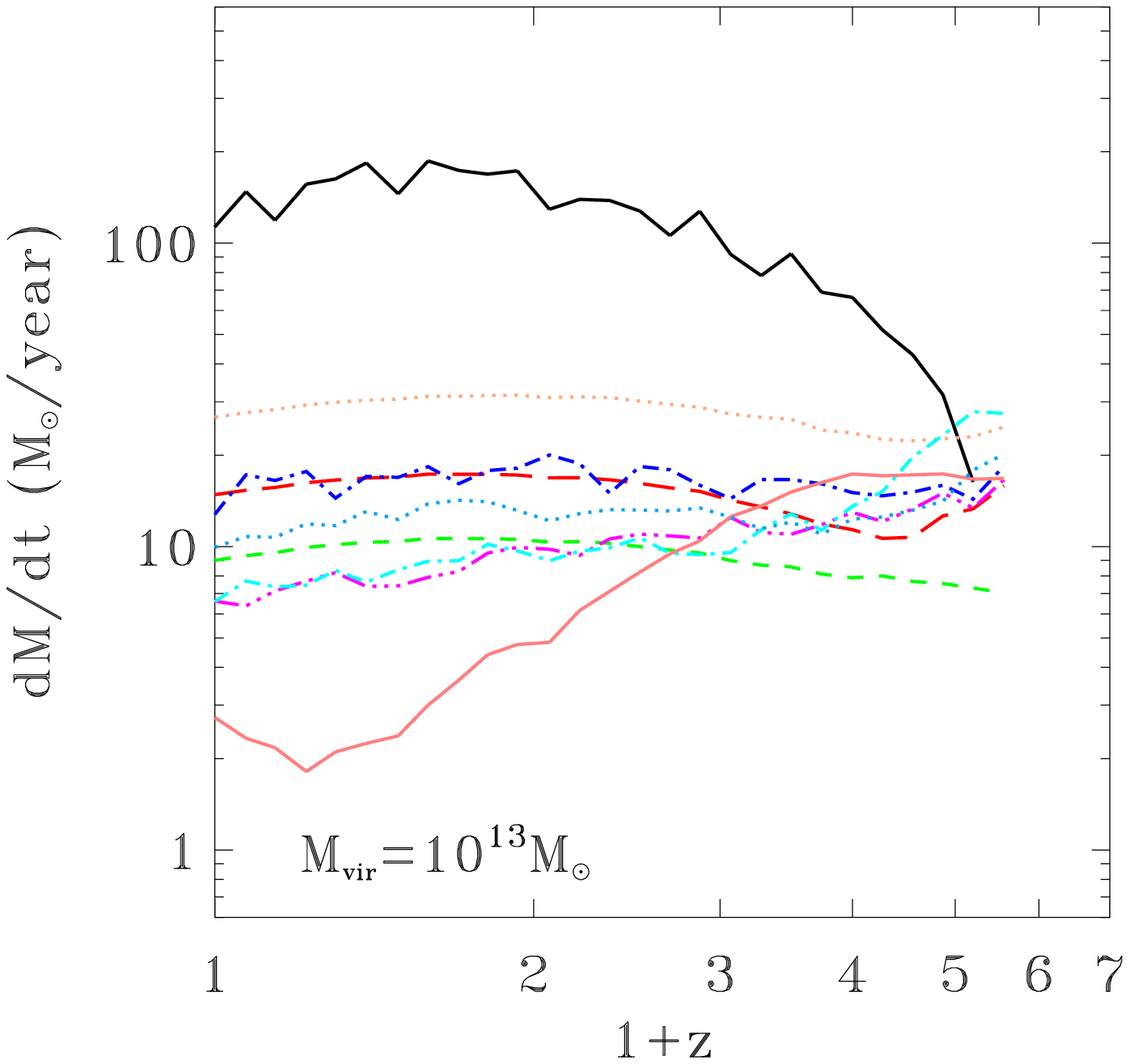, scale=0.5}
    \end{center}
  \end{minipage}
  \hfill
  \begin{minipage}[t]{.45\textwidth}
    \begin{center}
      \epsfig{file=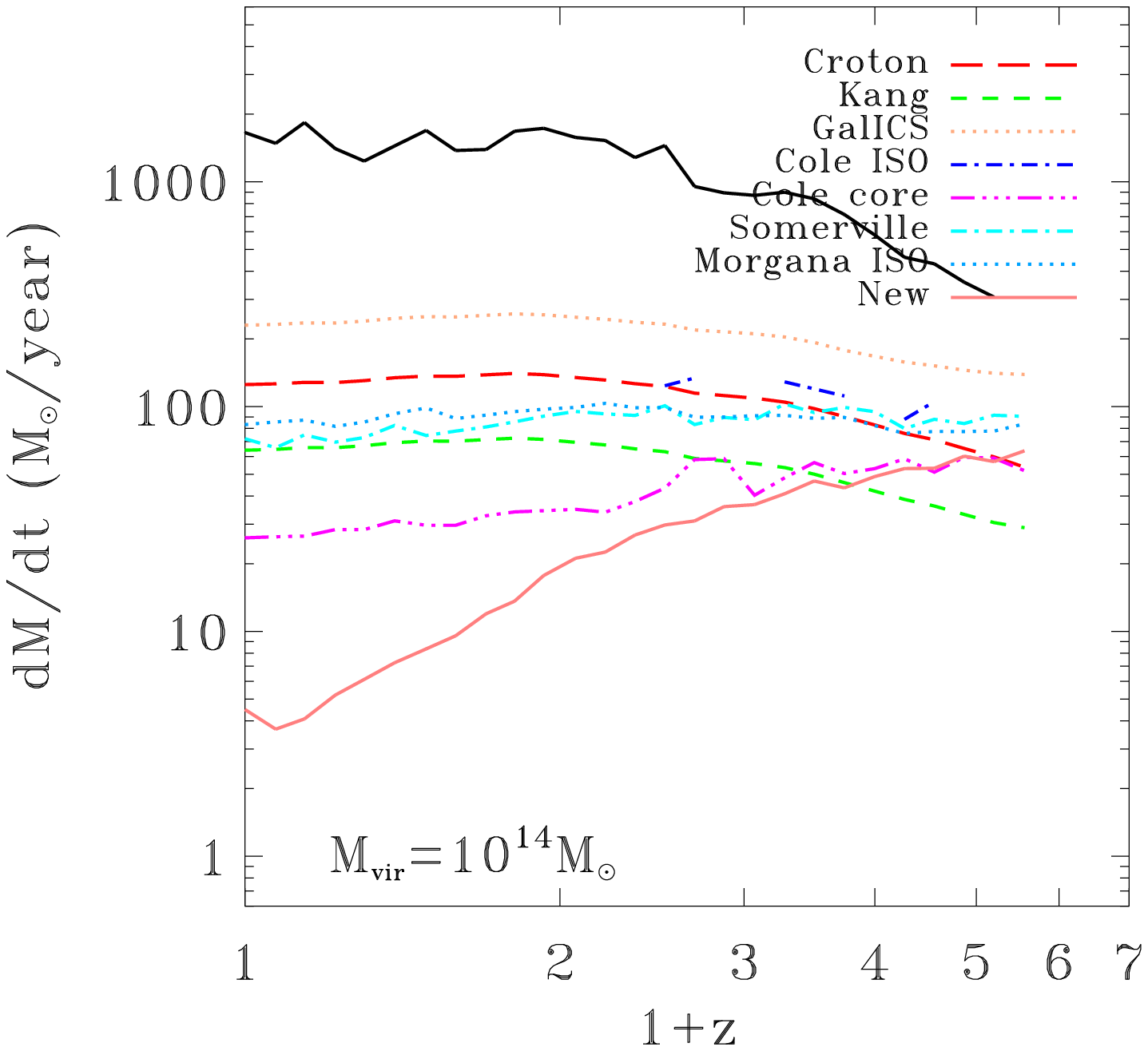, scale=0.5}
    \end{center}
  \end{minipage}
  \hfill
\caption{\large
The cooling rate as a function of redshift for the Monte Carlo 
generated merger tree SAMs.  The panels use the same final halo masses
as the previous figures.
%Each panel corresponds to a different final
%halo mass ($10^{11}, 10^{12}, 10^{13}$ and $10^{14}\msun$ as labeled).
For each halo mass, 100 merger trees are generated and
the results are averaged.
Only the cooling within the main branch haloes is shown.
The solid line denotes the average
baryon accretion rate of the haloes. The other lines
correspond to the cooling rates predicted by the different
cooling models as shown in the legend.
}\label{fig:mcr}
\end{figure}

\begin{figure}
  \hfill
  \begin{minipage}[t]{.45\textwidth}
    \begin{center}
      \epsfig{file=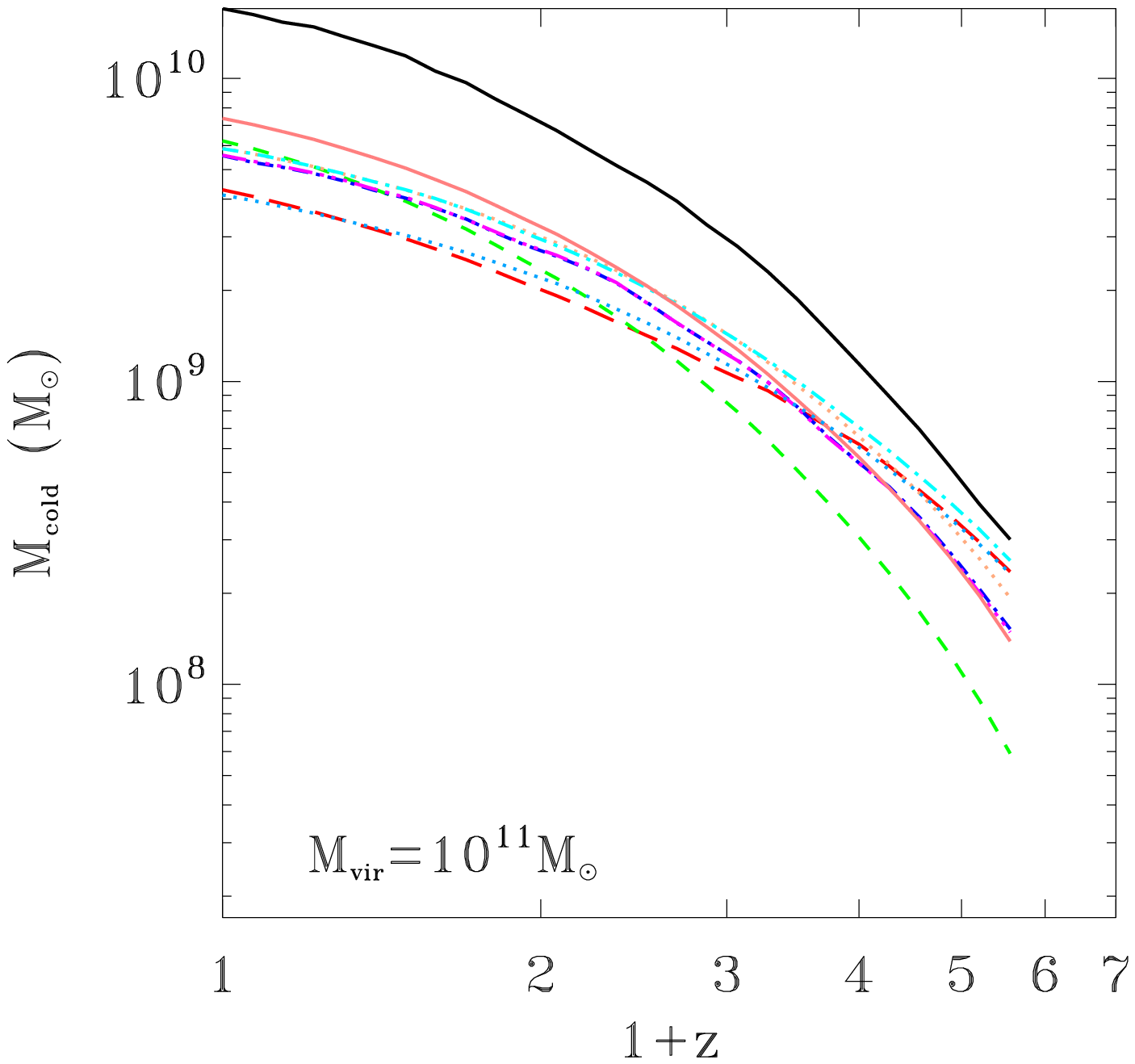, scale=0.5}
    \end{center}
  \end{minipage}
  \hfill
  \begin{minipage}[t]{.45\textwidth}
    \begin{center}
      \epsfig{file=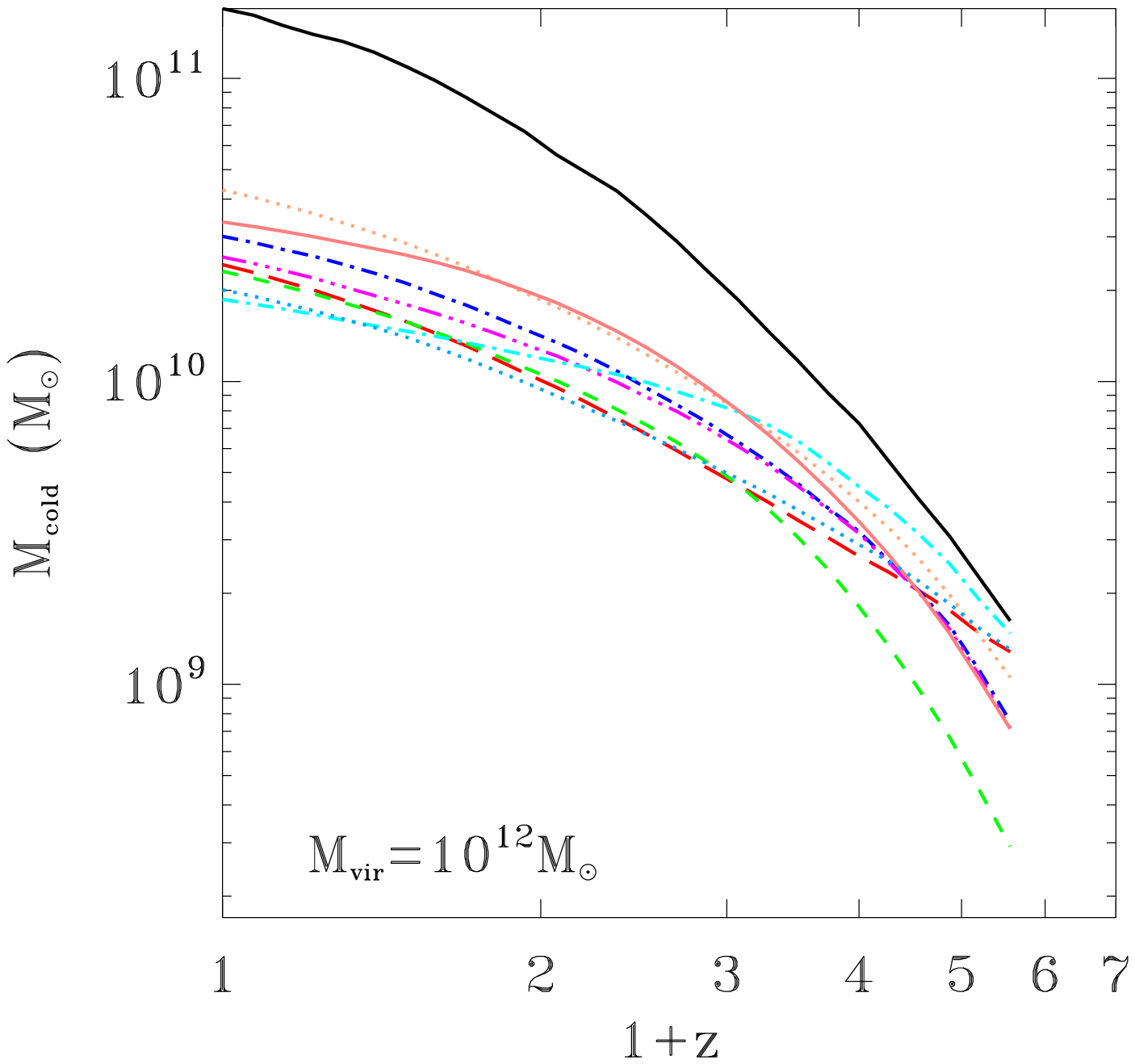, scale=0.5}
    \end{center}
  \end{minipage}
  \hfill

 \vfill
  \hfill
  \begin{minipage}[t]{.45\textwidth}
    \begin{center}
      \epsfig{file=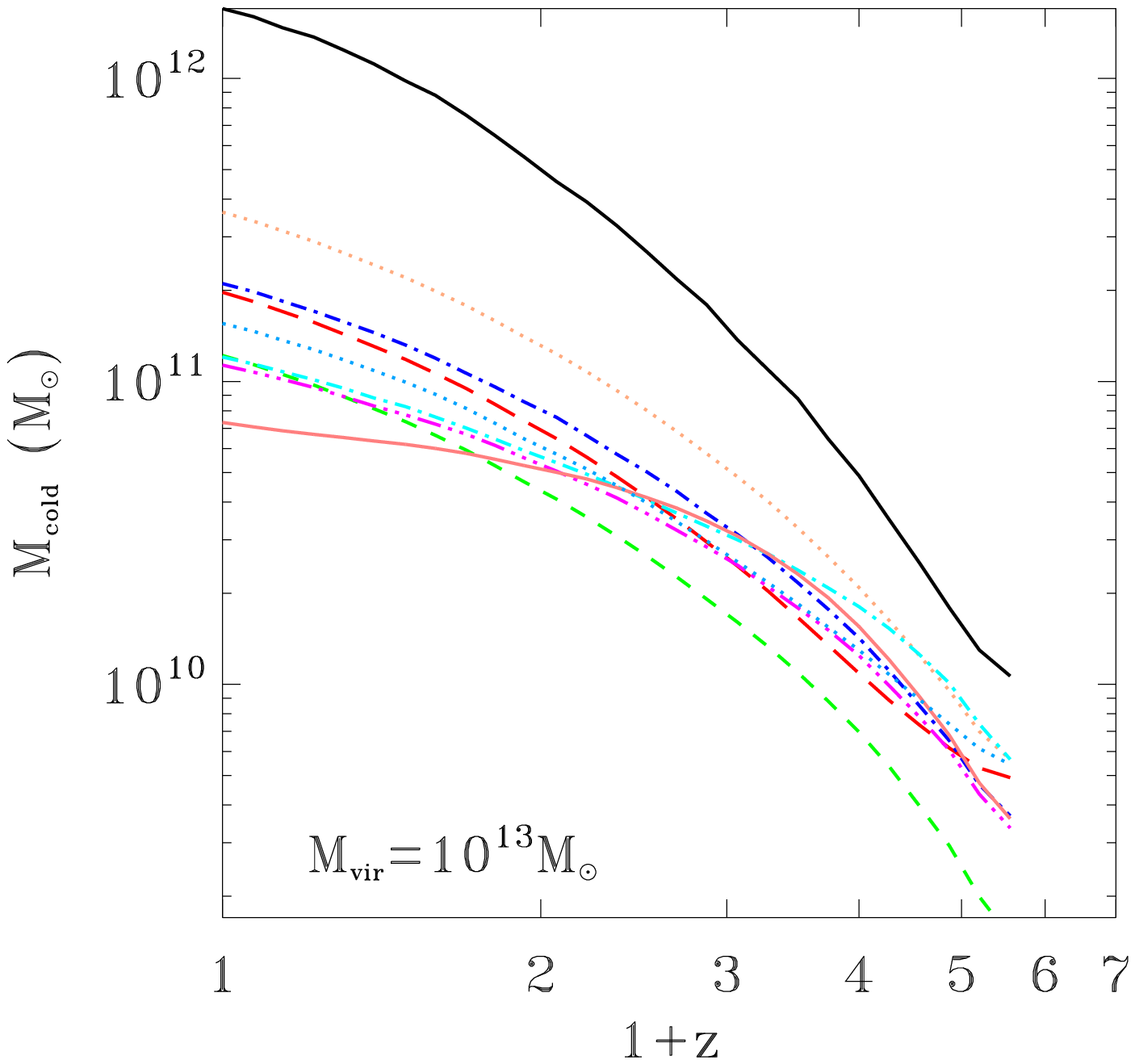, scale=0.5}
    \end{center}
  \end{minipage}
  \hfill
  \begin{minipage}[t]{.45\textwidth}
    \begin{center}
      \epsfig{file=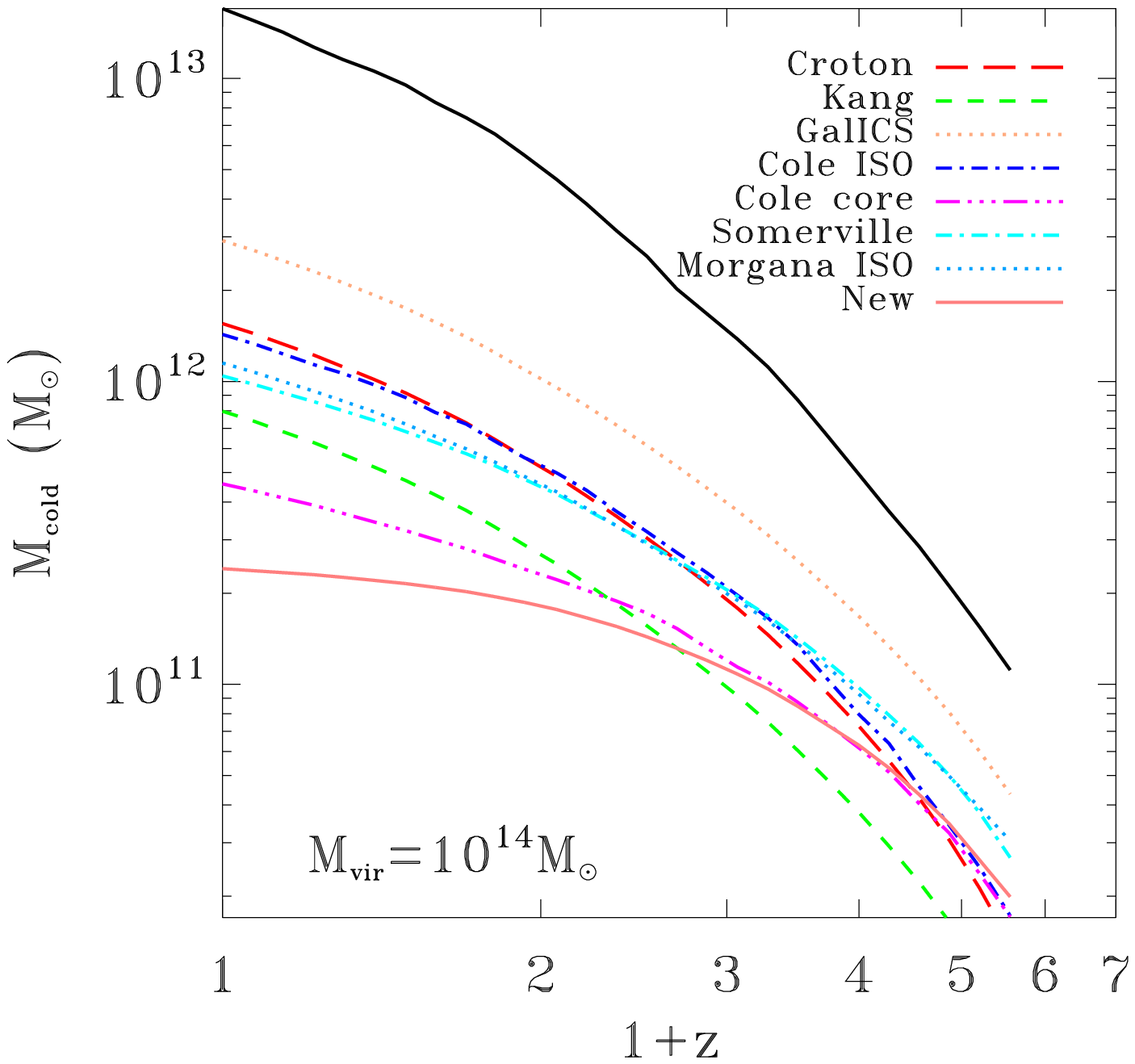, scale=0.5}
    \end{center}
  \end{minipage}
  \hfill
\caption{\large The cumulative
cold gas mass of the central galaxies for the
merger model in Figure~\ref{fig:mcr}
as a function of redshift. 
Only the gas cooled in the main branch is shown.}\label{fig:mcm}
\end{figure}

We illustrate the cooling rates and cold gas masses (both being the average
of 100 merger trees) of the main progenitors for the four halo masses
in Figures \ref{fig:mcr} and \ref{fig:mcm}.
Unlike the models with smooth MAHs, the central galaxies in
the merger tree models could have a considerable fraction of 
their cold gas mass acquired through merging satellite galaxies.
To exclude any differences that could result from the different treatment of
galaxy mergers in the SAMs and to concentrate our study on the cooling
of the halo gas, we exclude all the cold gas mass acquired through
mergers from our analysis. Even with this treatment, gas cooling
in progenitors can still have an important effect because it
lowers the amount of gas available for cooling in the descendant haloes.
Remember that the cumulative cold gas mass shown in Figure \ref{fig:mcm} represents
the result of cooling that occurs only in the main branch. 
 
The figures show some of the same features seen
in the simple smooth-accretion model,
but some new features arise owing to the merging nature of the
dark-matter haloes. First, the amounts of cumulative cold gas
shown 
in Figure \ref{fig:mcm} are smaller than the corresponding results 
shown in Figure \ref{fig:scm}, owing to the excluded gas that cooled in merging progenitors.
Second, the cooling rate in the Croton model remains
higher than in the Kang model, but for small haloes the differences
between the two models are reduced since the Croton model switches to
``slow'' cooling at earlier times than the Kang model.
This occurs for small haloes because the switch
between cooling regimes almost always occurs at a time when
the halo mass is about $3\times10^{10}\msun$ for the Croton model
and about $2\times10^{11}\msun$ for the Kang model, as shown
in Figure \ref{fig:scm}. For the lowest mass haloes ($10^{11}\msun$), 
the Croton model predicts a higher cooling rate at high redshift, 
but this leaves less hot gas to cool at lower redshift. In contrast, the
Kang model leaves more gas to cool at low redshift, 
leading to higher cooling rates at late times in these low mass haloes.
Third, the MORGANA model for an isothermal profile predicts 
lower cooling rates than the Kang model 
for the hierarchically growing halos. This seems contradictory to what we find in the 
case of smooth-accretion halos, where the cooling rates predicted by 
the MORGANA model are higher than the Kang model. 
A similar behaviour is also shown in \citet{DeLucia2010}, in which 
the authors compared the MORGANA model with the Croton model and the Durham 
model and, against their expectations, they found that
the MORGANA model does not predict more efficient cooling than the Croton model.
This result seems surprising,
but it is easy to understand when the hierarchical 
formation of dark matter halos is taken into account. As we have shown, the 
MORGANA model cools faster at early times in progenitor halos,
the main branch halo at later times has less hot gas to cool and, therefore,
has a lower cooling rate. 
Fourth, adding a central core 
to the gas density distribution in the Cole model yields lower cooling 
rates for massive haloes, yet it does not have a significant effect 
on low-mass haloes because the cooling in these haloes is so efficient 
that most of the gas can cool even if the gas density is lowered by 
the presence of a core. Finally, the GalICS model in general predicts 
cooling rates that are higher than any of the other models.

\begin{figure}
  \hfill
  \begin{minipage}[t]{.45\textwidth}
    \begin{center}
      \epsfig{file=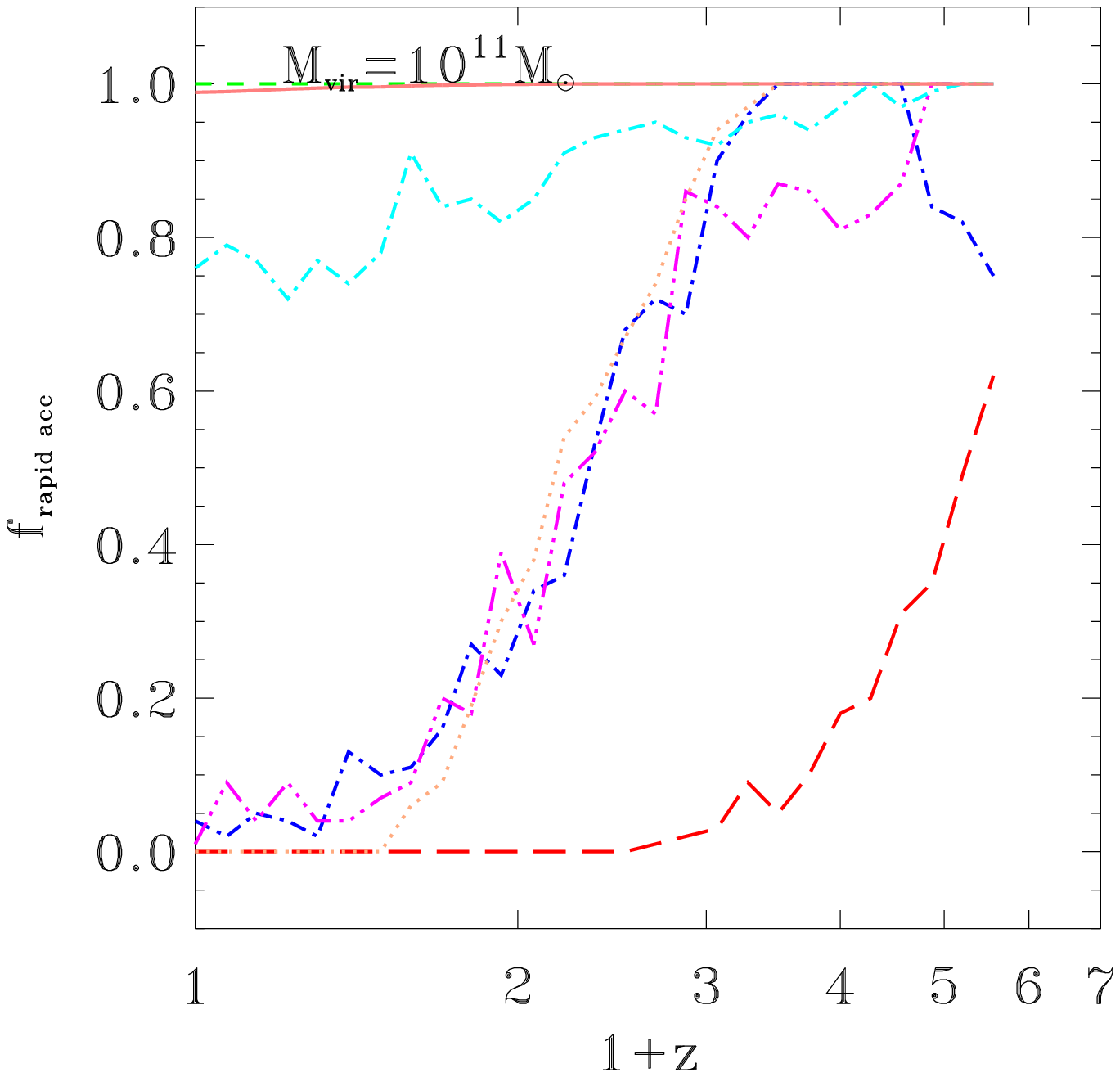, scale=0.5}
    \end{center}
  \end{minipage}
  \hfill
  \begin{minipage}[t]{.45\textwidth}
    \begin{center}
      \epsfig{file=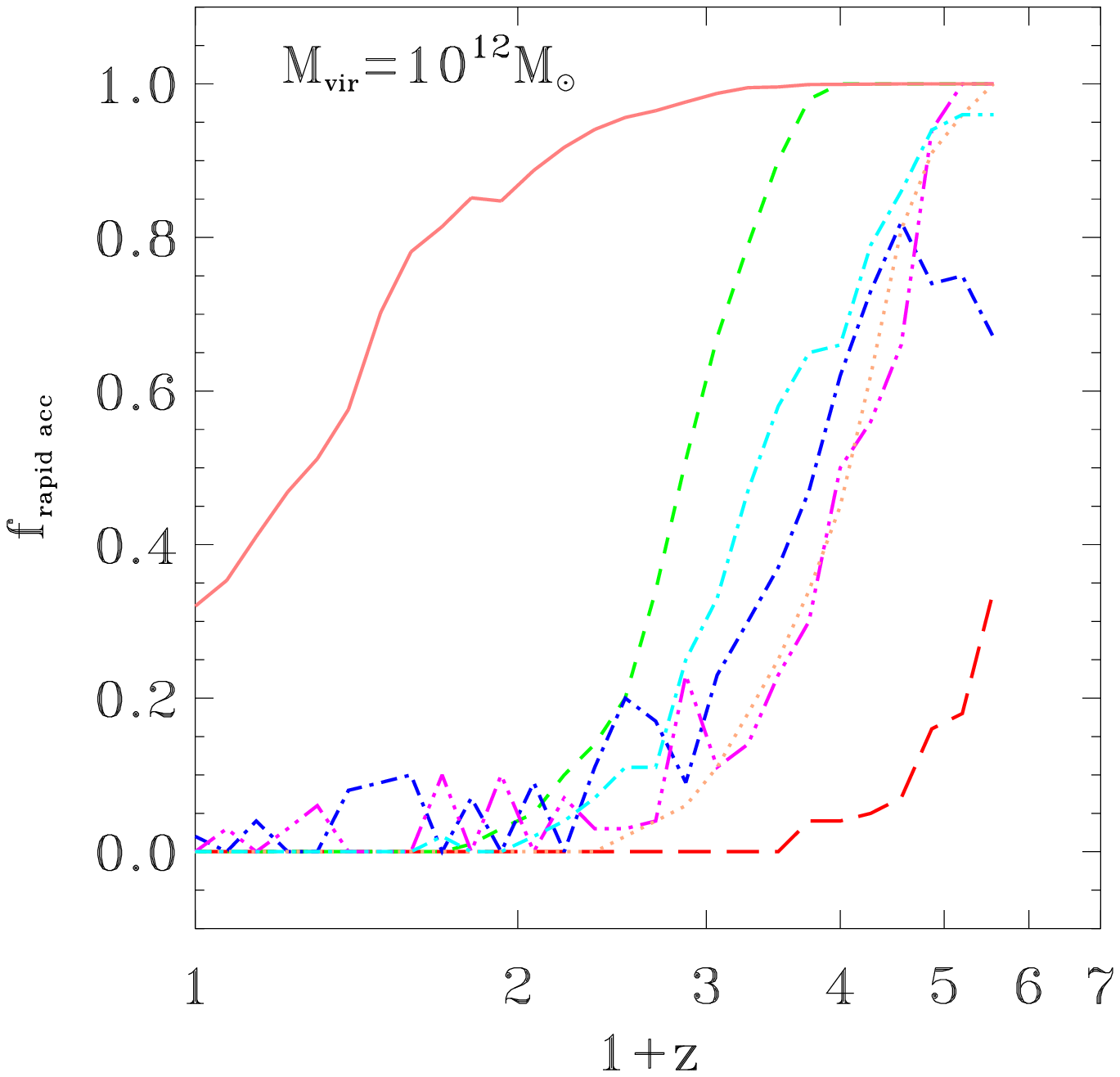, scale=0.5}
    \end{center}
  \end{minipage}
  \hfill

 \vfill
  \hfill
  \begin{minipage}[t]{.45\textwidth}
    \begin{center}
      \epsfig{file=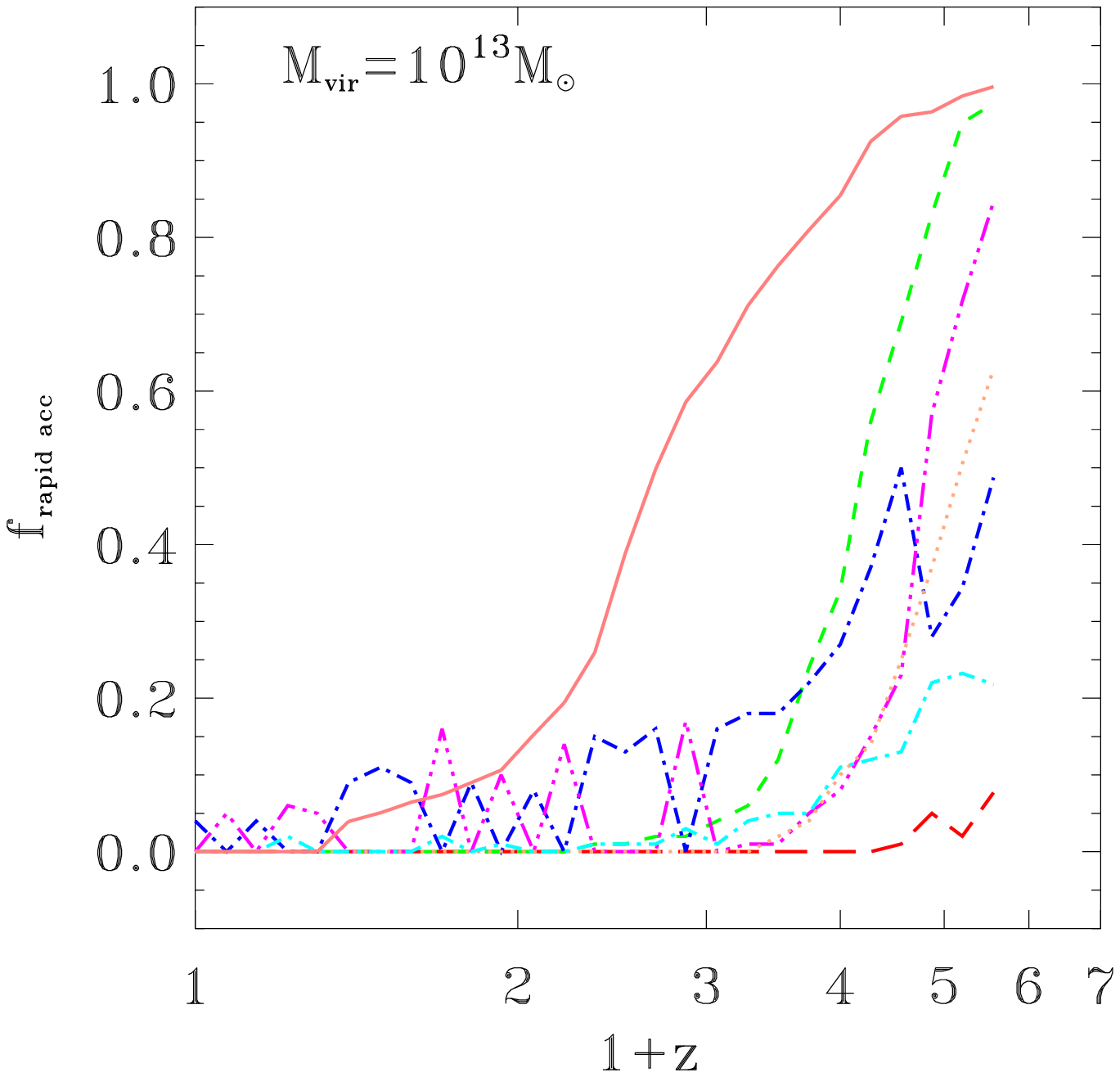, scale=0.5}
    \end{center}
  \end{minipage}
  \hfill
  \begin{minipage}[t]{.45\textwidth}
    \begin{center}
      \epsfig{file=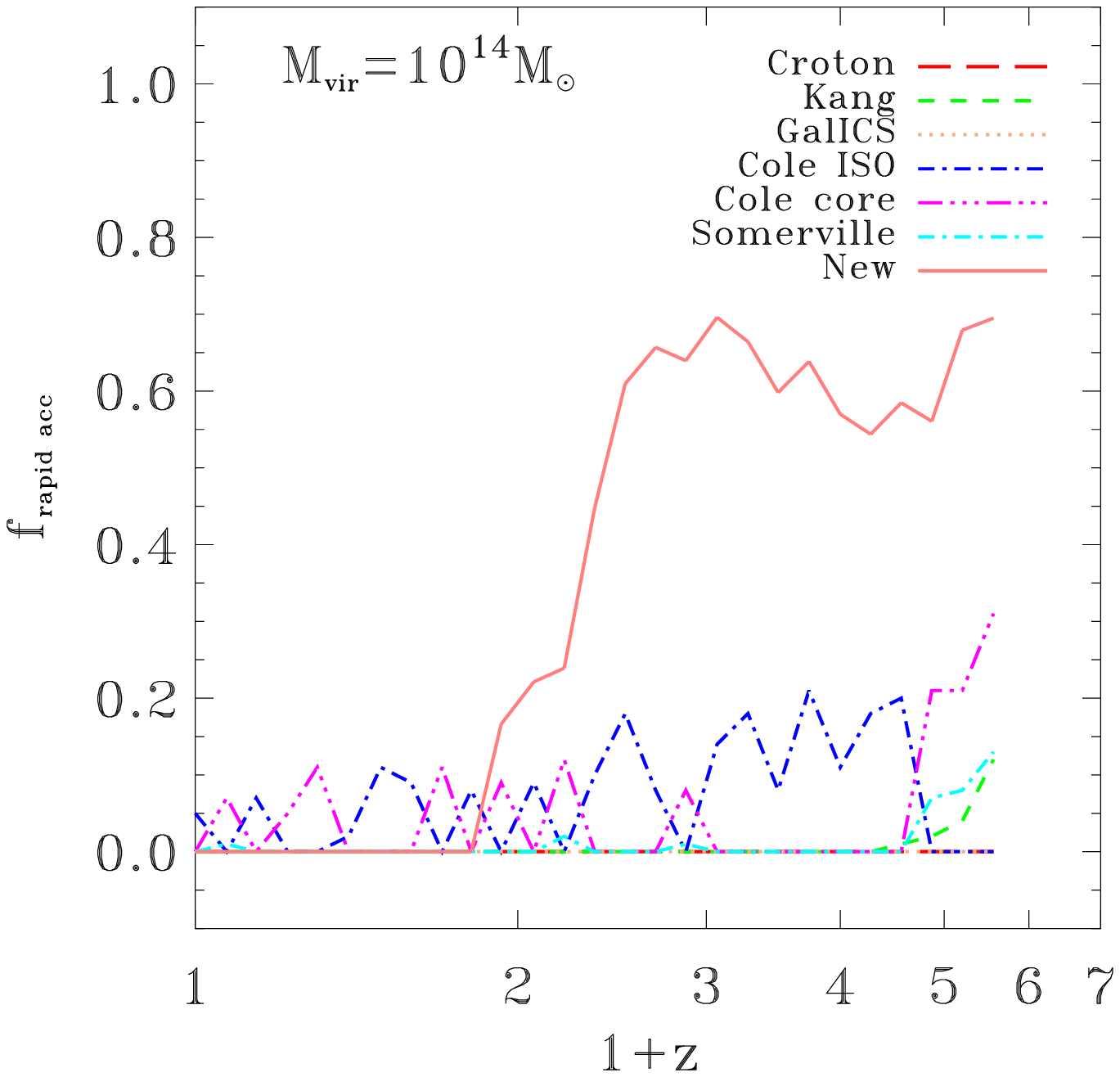, scale=0.5}
    \end{center}
  \end{minipage}
  \hfill
\caption{\large
The fraction of main branch haloes in the ``rapid'' accretion mode
as a function of redshift for the merger model
used in Figure~\ref{fig:mcr}.
}\label{fig:mcf}
\end{figure}

In the SAMs considered here, except the MORGANA model,  
a halo is either in the ``rapid'' or 
the ``slow'' cooling mode. However, haloes with the same final mass may switch between 
the modes at different times owing to their individual merging history.
Figure \ref{fig:mcf} show the fraction of haloes in the ``rapid'' 
mode as a function of redshift for haloes of different final masses, 
as predicted by the cooling models. This fraction is determined from 
a limited halo sample at each redshift snapshot and is somewhat 
noisy. Nevertheless, several features can be seen 
from the figure. For small haloes ``rapid'' mode accretion 
dominates during the entire formation history in the Kang model and the
Somerville model, while the other models switch to ``slow'' mode
at $z\sim2$. The intermediate mass haloes all have transition from
the ``rapid'' mode to the ``slow'' mode at some time during
their formation histories but the transition occurs at different
times in the different models. The transition times predicted
by the Croton model and Kang models are consistent with the
prediction of the simple smooth accretion model presented
above, but the GalICS model and the two Cole models predict transition 
times that are different from the smooth accretion model.  As discussed 
in \S 2, the definition of ``rapid'' and ``slow'' modes for the GalICS model
depends on the time step,  which differs between the smooth-accretion and
merger tree models, and so the difference in the predicted transition
times is not surprising. In the Cole models, since the gas distribution
is reset whenever the halo doubles its mass, the results can be
sensitive to the details of the halo assembly history. Finally, for massive 
haloes, all the models predict that ``slow'' mode accretion dominates 
through their entire formation history. The Croton model has 
the smallest fraction of haloes in the ``rapid'' mode at any redshift 
and mass, making this accretion mode the least important in their model.
However, models that do not renormalise the gas density profile at every 
time step often have some small fraction of haloes in the ``rapid'' mode 
even at late times and for massive haloes.

\begin{figure}
  \hfill
  \begin{minipage}[t]{.45\textwidth}
    \begin{center}
      \epsfig{file=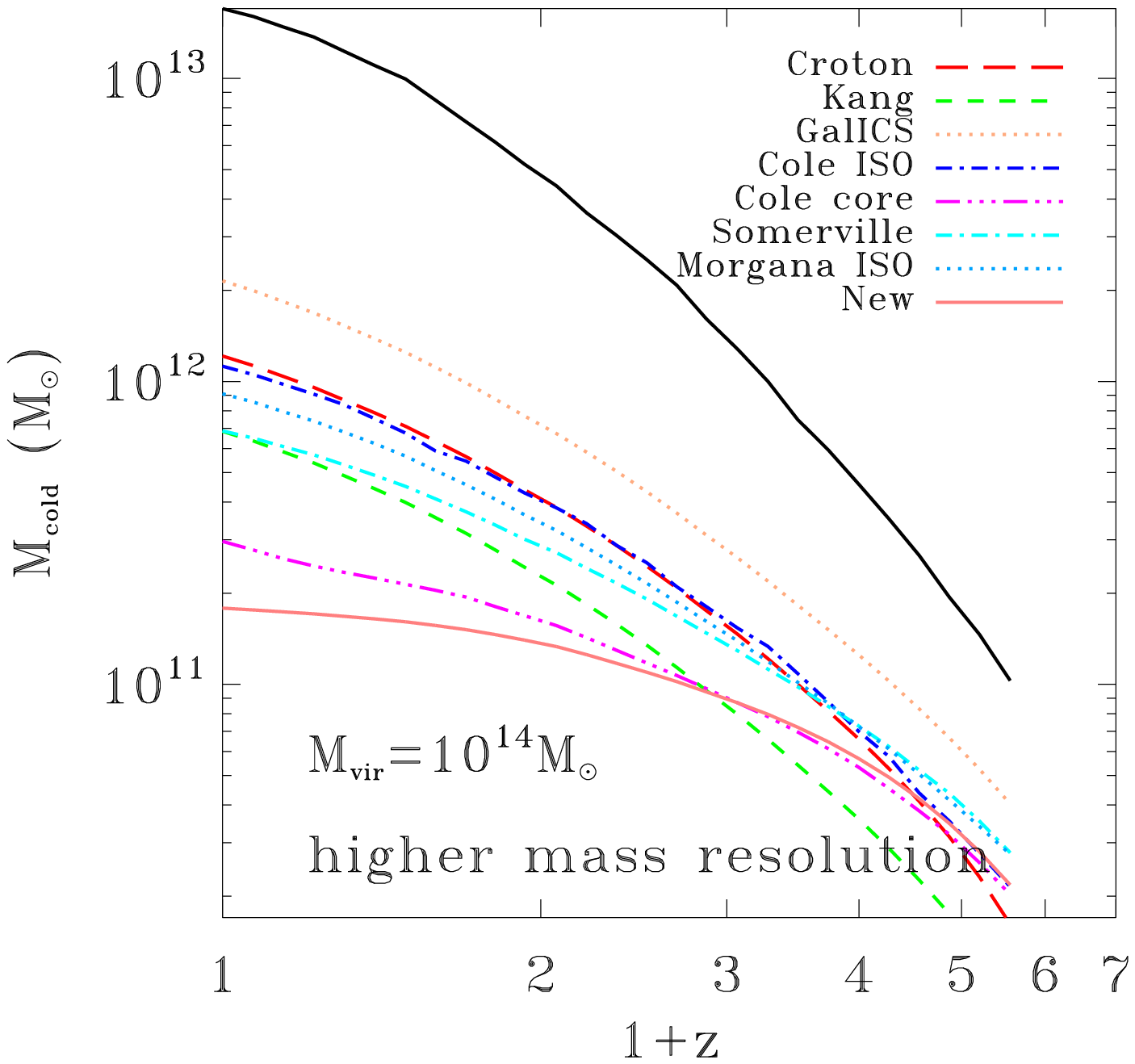, scale=0.5}
    \end{center}
  \end{minipage}
  \hfill
  \begin{minipage}[t]{.45\textwidth}
    \begin{center}
      \epsfig{file=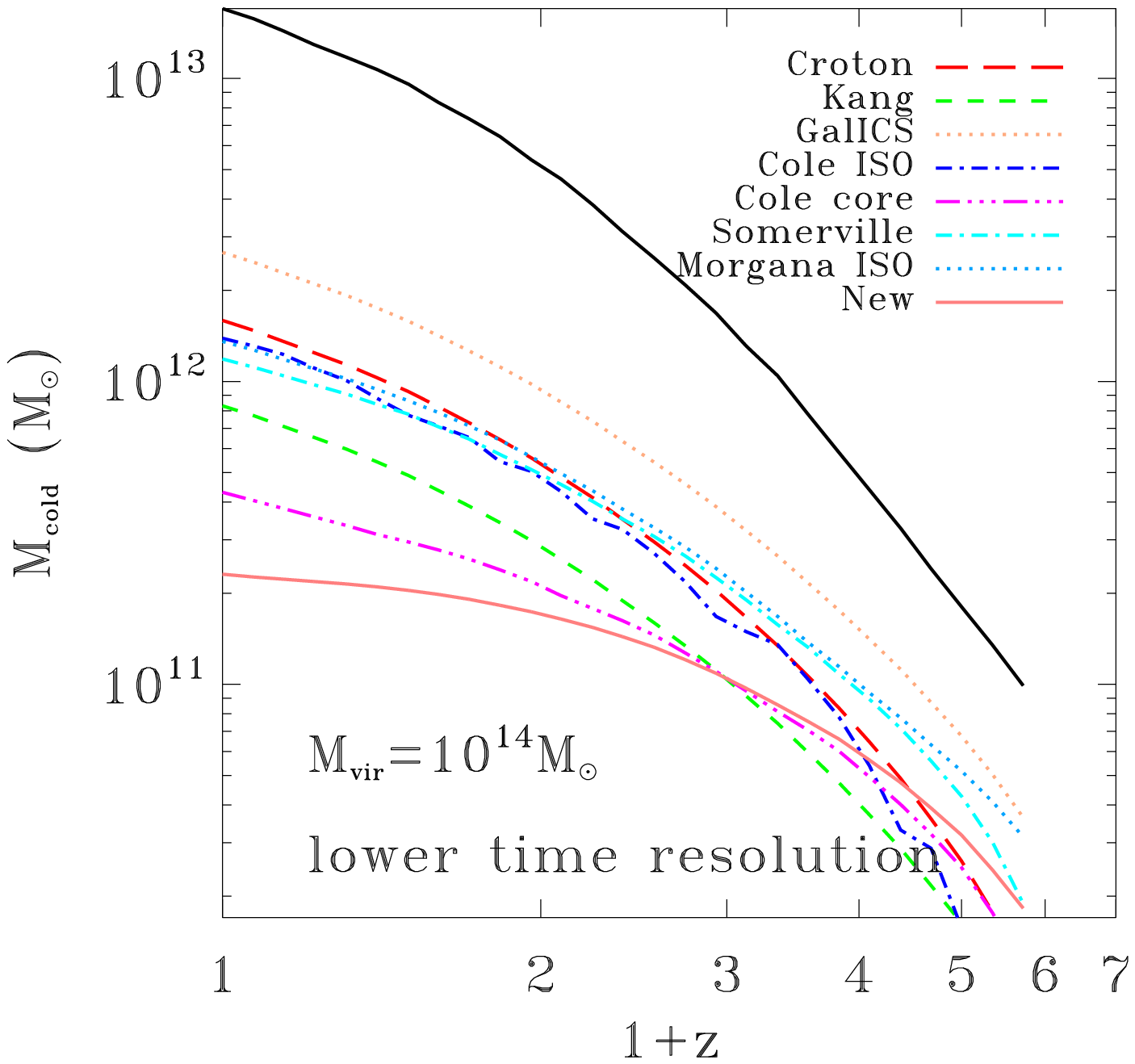, scale=0.5}
    \end{center}
  \end{minipage}
  \hfill
\caption{\large
Mass and time resolution tests of SAMs shown as the cumulative 
cold gas mass of $10^{14}\msun$ halos. 
The left panel shows the predictions with a merger tree 
mass resolution five times higher than that for the lower
right panel of Fig.\ref{fig:mcm}. 
The right panel shows the predictions using merger tree time steps 
that are two times larger. 
}\label{fig:rt}
\end{figure}

The mass resolution adopted for these merger trees is 0.001 of the final 
halo mass. For the two lower masses, this choice 
achieves higher mass resolution than those adopted in the literature. For the 
$10^{13}\Msun$ halos, the resolution is similar to that of the models based 
on the Millennium simulation \citep{Springel2006}. The mass resolution 
for the $10^{14}\Msun$ halos is lower than what is typically used in SAMs. 
We notice that, for some models at very high redshift, 
the predicted cold gas mass is not 
much higher than the adopted mass resolution, which suggests that these 
results are likely affected by the finite resolution of our model.
To test the effect of mass resolution on our results we generate the 
same number of merger trees with a final mass of $10^{14}\Msun$ using 
a mass resolution  of $2\times 10^{10}\Msun$. This mass resolution is 5 times higher 
than the one used in our previous discussion. We keep everything else
the same and run the models on the merger trees. The left panel of 
Figure \ref{fig:rt} shows the predicted cold gas of the central galaxies as a function of redshift. 
We find that increasing the mass resolution has no effect on the accretion
histories at early times, indicating that the mass resolution we adopt
can well resolve the accretion of these halos even at early times when
the halo mass is low. At late times, however, the predicted cold gas mass
shows a mild deviation from the fiducial models. The high resolution models
predict lower cold gas mass. In the most extreme case, the GalICS model,
the cold gas mass is reduced by 1/3, and the decrease is smaller for the other 
models.
Although the decreases in the cold gas mass 
at late times are not significant, the effect is systematic. 
This occurs because with a higher mass resolution,  more gas mass cools 
in progenitors instead of in the main branch halos and hence less gas mass is 
available to cool in the main branch halos at later times.
The relative behaviour between the models, however, stays the same,
confirming our previous discussion.
We also test the effect of the time resolution of the merger trees. 
For the $10^{14}\Msun$ halos, we use the same merger trees, but only 
save 30 snapshots between $z=0$ and 6, instead of 60, so that 
the time resolution 
is reduced by a factor of 2. We run the models on the reduced trees and 
show the results in the right panel of Figure \ref{fig:rt}. 
The predicted cold gas mass is stable and not sensitive 
to the choice of time steps, except at very early times, $z > 4$.

%%%%%%%%%%%%%%%%%%%%%%%%%%%%%%%%%%%%%%%%%%%%%%%%%%%%%%%%%%%%%%%%%%%

\subsection{Comparison to Numerical Simulations}\label{subsec:simu}

In this section, we compare the SAM results for cooling in merging haloes with
a large-box cosmological Smoothed Particle Hydrodynamics (SPH) simulation \citep{Kerevs2009}.
The simulation models a 50$\hmpc$ comoving, periodic cube using $288^3$ dark
matter and $288^3$ gas particles, i.e. around 50 million particles in total.
The mass of each gas and dark matter particle is $9.0\times 10^7 \Msun$ 
and $4.4 \times 10^8$, respectively.  Hence, $\sim 10^{11}\Msun$ halos in
the simulation are typically composed of about 200 gas and 180 dark matter
particles.  There are a larger number of gas particles because
the baryon fraction has about a 10\% excess over the cosmic average value
for small halos, but converges to the cosmic baryon fraction for massive halos 
\citep{Keres2007,Faucher-Giguere2011}. 
However, since haloes of these small masses accrete their gas through cold
mode accretion, to recover the converged gas accretion rates it is only 
necessary to correctly reproduce the gravitational gas accretion rate,
something that is easily accomplished using this many particles.
Gravitational forces are softened using a cubic spline kernel of comoving
radius $10\kpch$, approximately equivalent to a Plummer force softening of
$7.2\kpch$.  The virial radius of the small halos ($\sim 10^{11} \Msun$) is
$121 \kpc$ at $z=0$, which means that the gravitational softening is about
1/12 of the virial radius.  For massive halos ($\sim 10^{13} \Msun$),
the softening is less than 1/50 of the virial radius at $z=0$.
The initial conditions are evolved using the SPH code, Gadget-2
\citep{Springel2005a}.  This code uses the entropy and energy conserving formulation
of SPH from \citet{Springel2002}, which avoids numerical problems with hot 
gas over-cooling in massive haloes \citep{Springel2002, Keres2007}.
Our version of Gadget-2 includes modifications to incorporate
gas cooling, a photoionising UV background, and star formation.
We include all the relevant cooling processes with primordial abundances as
in \citet{Katz1996}, similar to the cooling implementation in the SAMs.
We do not include any metal enrichment or cooling processes associated
with heavy elements or molecular hydrogen. This simulation also includes 
a spatially uniform, extragalactic UV background that heats and ionises the gas.
The background flux first becomes nonzero at $z=9$ and is
based on \citet{Haardt2001}; for more details see \citet{Oppenheimer2006}. 
The UV background pre-heats the gas and modifies the cooling curve, 
but this change affects only galaxies far below the resolution limit 
of the simulations at early times, growing to around the resolution limit 
only at late times. 
Our tests show that incorporating the UV background 
and the cooling function used in the simulation do not significantly affect
the accretion rates and the mass accumulation history of SAMs presented 
in this paper, although the transition from rapid to slow cooling regime 
can be affected in some models \citep[see][]{Kerevs2005}. 
When the UV background is included, cooling in small progenitor halos is suppressed 
and, therefore, the more massive descendant halos will have more hot gas.  
As a result, the addition of a UV background results in slightly higher 
accretion rates. Our tests show that the accretion rates of 
the $\sim 10^{11}\Msun$ halos in SAMs are enhanced by less than 20\% when 
the UV background is included. For more massive halos, the effect becomes even smaller.
The simulation also 
includes the star formation prescription and the sub-resolution two-phase 
medium, which is pressurised by supernovae, following the formalism of
\citet{Springel2003}. However, since supernova winds are not explicitly included in
this simulation, star formation and the supernova pressurised two-phase
medium does not affect any halo gas, i.e.  it only affects gas within the
galaxies themselves, and hence is irrelevant for the comparisons we present here.
Since the simulation and the SAMs differ in their star formation modelling,
we compare them using the total baryonic masses of the galaxies, 
adding together the stellar and cold gas components.

To identify bound groups of cold, dense baryonic
particles and stars, which we associate with galaxies,
we use the program SKID\footnote{
\tt http://www-hpcc.astro.washington.edu/tools/skid.html}
(see \citealp{Kerevs2005} for more details).
Briefly, a galaxy identified by SKID contains bound stars and gas with an
over-density $\rho/\bar{\rho}_{\rm baryon} > 1000$
and temperature $T < \mbox{30,000 K}$. Here, we slightly modify
this criterion and apply a higher temperature threshold at
densities where the two-phase medium develops.  Such a modification
is necessary to allow star forming two-phase medium particles
to be a part of a SKID group,  since at high densities the mass-weighted
temperature in the two phase medium can be much higher than 30,000~K.
To identify haloes, we use Spherical Overdensity (SO) algorithms with the algorithm
and virial overdensity as in \citet{Kerevs2005}, adjusted to the new cosmology.
We define ``central'' galaxies simply by choosing the
most massive galaxy in each halo, regardless of position.

To avoid counting the accretion of sub-resolved groups as smooth
accretion, we define smooth gas accretion as the accretion of gas
particles that were not part of any SKID identified group at the previous
time. In practise, this procedure will avoid counting sub-resolution 
mergers down to galaxies with $\sim 20$--30 particles, 
i.e.\ baryonic galaxy masses of $\sim 2\times 10^{9} \msun$,
approximately the mass where the galaxy mass function in our simulation
drops rapidly owing to our limited resolution. Readers are referred 
to \citet{Kerevs2009} for a more detailed description of the simulation.
There, the authors discuss that a large fraction of the cold-mode 
accretion in haloes more massive than $\sim 5 \times 10^{12}\msun$
at late redshifts is contributed by ``cold drizzle'',
i.e.\ accretion of individual cold particles or small groups of
particles below the resolution limit that might be a consequence of
poorly understood numerical effects. In this comparison, we still keep 
the contamination of ``drizzle'' as there is no unique way to remove it. 

To make a fair comparison between the SAMs and the simulation, we generate a set of
merger trees with the same cosmological model, the same volume, and the same 
mass resolution as the simulation using the Monte-Carlo method. We then
run all the SAM cooling models for these halo merger trees to predict the
cooling rates and the total cold gas masses. A comparison between the
simulation and the SAM using merger trees from the simulation itself
might be more direct,
but for the purposes of this study, the Monte-Carlo merger trees are sufficient and
better facilitate comparisons with the previous sections.
We first draw a random sample
of dark matter haloes using the EPS halo mass function from a
$(50\hmpc)^3$ volume for the same cosmological model as the simulation.
We only generate the merger trees for haloes with virial masses larger than
$5\times 10^{10}\msun$ at $z=0$ for the sample, and we set the halo virial
mass resolution to be $2\times 10^{10} \msun$ to be
approximately consistent with the effective resolution of the simulation.
The time step of the merger tree is chosen to be the same as 
that used in \S\ref{subsec:mhalo}. 
We apply every SAM cooling model to the merger trees
and compare the SAM predictions with the simulation.

To avoid the complication of tracing the main branch of the
halo merger trees in the simulation, unlike in the previous sections,
we look at haloes in a {\em fixed} mass range at different redshifts.
At each SAM or simulation snapshot, we select haloes in the mass ranges
$10^{11\pm 0.15}\msun$, $10^{12\pm 0.15}\msun$, and $10^{13\pm 0.20}\msun$.
We then calculate the cold gas accretion rate for central galaxies 
in these haloes and the total halo gas mass that has not collapsed onto 
any galaxies, centrals or satellites, in the haloes for both the simulation 
and the SAMs. For the simulation, we choose four snapshots at $z=0$, 
1, 2, and 3, and determine the accretion rates in each of these 
snapshots using the change of the cold gas mass between $z$ and 
$z+0.1$. For the SAMs, we use 15 snapshots
evenly distributed in $\log(1+z)$ from $z=6$ to $z=0$.

\begin{figure}
\epsfig{file=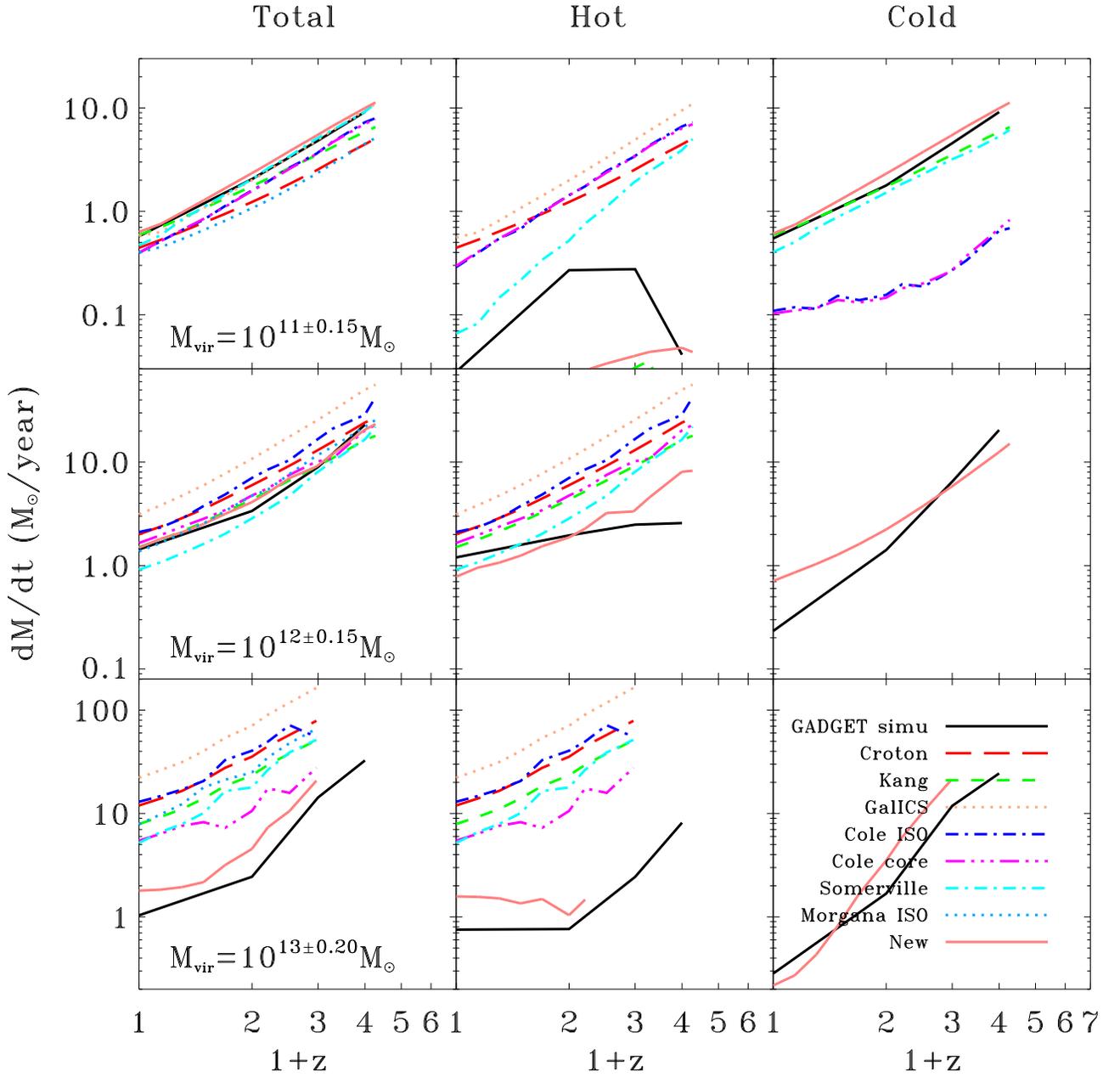}
\caption{\large The average gas accretion rate onto the central objects
of haloes with different halo mass ranges noted in the
left panels as a function of redshift. The first column shows the total
accretion rate; the second column shows the ``hot'' accretion rate;
and the third column shows the ``cold'' accretion rate. The solid
black lines show results from the SPH simulation, while the colour coded
lines show the predictions of the SAM cooling models. 
Halo mass range is held constant with redshift.
}
\label{fig:simu-sam}
\end{figure}

For each halo mass range, we plot the averaged cold gas accretion
rate of the central galaxies as a function of
redshift in each row of Figure \ref{fig:simu-sam}.
For the simulation, each of the three columns shows the total
accretion rate, the accretion rate in the hot mode
and the accretion rate in the cold mode, respectively. 
As in the previous section, we do not include any cold
gas from merging satellite galaxies in the SAMs. Correspondingly, 
in the simulation, we define the gas accretion as the 
accumulation of gas that was outside of any galaxy at the previous simulation output.  
We show the cooling rate from the hot halo gas in the simulation
in the middle column. Since haloes may accrete cold gas directly,
any cold gas that collapses onto the central
galaxies but is not associated with merging galaxies is accounted as
cold-mode accretion and plotted in the last column.
The SAMs do not explicitly model the cold halo gas component,
so the entire gas accretion is simply defined as
the gas mass that cools from the hot halo and collapses onto the central
galaxy. However, it is generally believed that the ``rapid'' and ``slow'' mode
cooling in SAMs are proxies for the cold and hot-mode
accretion observed in simulations. Following this idea, we split the accretion
in SAMs according to the definition of the ``rapid'' and ``slow'' cooling mode
for each of the models and plot them in the corresponding columns to compare with the simulation.

These plots illustrate several interesting points.
We find that the SAMs have considerable differences in their predicted cooling rates,
both in total and in the different modes, following the
patterns seen in previous sections. For $\sim10^{13}\msun$ haloes, the predicted
accretion rate differs by up to a factor of 4. Compared with the simulation, 
we find that none of the SAMs predicts cooling rates consistent with the simulation
for all halo masses. For $10^{11} \msun$ haloes, where the cooling
is efficient at all redshifts, the cooling rate is strongly limited by
the halo baryon accretion rate which is the same for all the SAMs, so the
predicted cooling rates do not differ very much.
However, the split between the accretion modes does not agree with
the simulation, where galaxies in small haloes accrete gas mainly through cold accretion,
which approximately follows the infall of baryons into haloes (see \citealp{Kerevs2005}).
In contrast, SAMs predict a much higher accretion rate through ``slow'' cooling.
As shown in the predictions of SAMs, the rate of ``slow'' mode cooling makes up for the cold-mode
accretion in small haloes, which occurs in the simulation but is missing in those models.
Moreover, it is clear that most of the SAMs, except the Somerville
and GalICS models, still under-predict the total accretion rate for small haloes even
though they predict much faster cooling of hot gas than the simulation.
For $10^{12} \msun$ haloes, the total accretion rates over redshifts predicted
by the SAMs show the same trend as the simulation, and the amplitudes
are close to the simulation within a factor of 2--3. However, the models 
attribute all the accretion to the ``slow'' accretion,
which results in zero ``rapid'' mode accretion and an over-prediction of hot
mode accretion at high redshift. For $10^{13} \msun$ haloes, 
the cooling rates predicted by the SAMs are 
at least 3 times higher than the simulation results, largely owing to 
the wrongly assumed steep gas density profile, and again none of the SAMs predict any
cold or ``rapid'' mode accretion in this halo mass bin.
(One should be cautious about the discrepancy in cold accretion
rates in massive haloes owing to the previously mentioned concerns
that the cold ``drizzle'' in simulated massive haloes might be 
purely of numerical origin.)
In summary, these results indicate that without modelling the cold halo gas, 
SAMs generally under-predict the cold-mode accretion and over-predict cooling in the hot
accretion phase.  As a result, the SAMs over-predict the
cooling at the high-mass end.

\begin{figure}
\epsfig{file=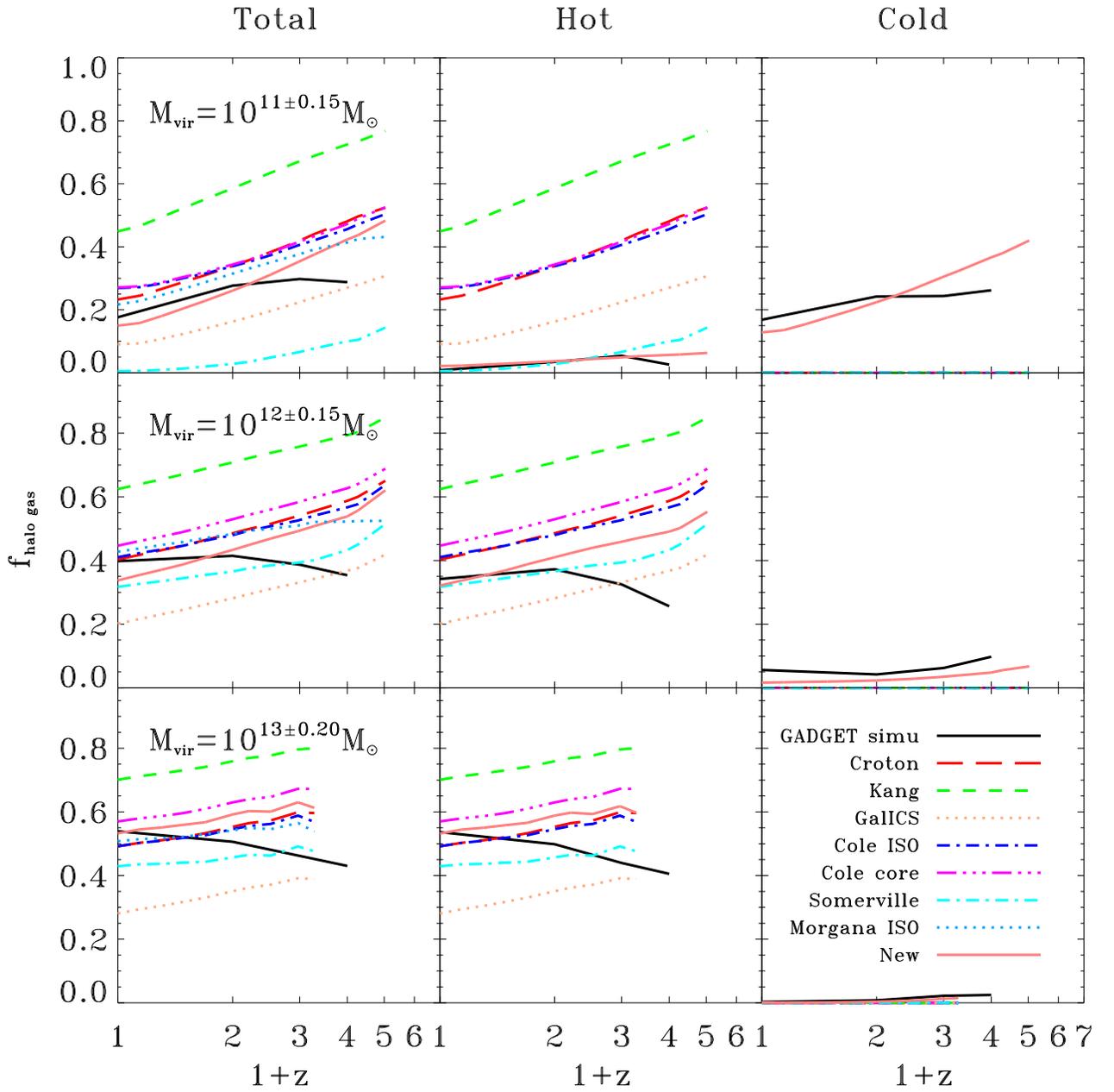}
\caption{\large The total uncollapsed, hot and cold halo gas fractions
as functions of redshift predicted by different SAMs compared with
fractions of the corresponding components in the simulation.
}\label{fig:simu-sam-hmf}
\end{figure}

The accumulated total baryonic mass of the central galaxy is difficult
to compare between simulations and SAMs, because much of this mass comes
from the merging of satellite galaxies. This happens naturally in
simulations but depends on the model recipe for dynamical friction
in the SAMs, including such parameters as the assigned Coulomb logarithm
and tidal stripping, models of which have large uncertainties \citep{Hopkins2010b, DeLucia2010}. 
To avoid these problems, we focus on the uncollapsed gas mass,
comprising both the total hot gas mass and the cold gas that is contained by a halo
but is not associated with any galaxies identified
by SKID.
We compare the averaged uncollapsed gas mass
fraction, i.e. uncollapsed halo gas mass divided by the total baryonic mass in the halo,
for each halo mass bin from all the simulation to the averaged total halo gas
mass fraction from the SAMs. The first column in Figure \ref{fig:simu-sam-hmf} plots 
this quantity versus redshift. Similarly, the second column shows
the hot halo gas mass fraction, and the third column shows the cold halo gas fraction.
None of the models predict any cold halo gas, simply because
this component is not included in any of the models.
The models show a large variation in their uncollapsed gas fraction.
Most of the models predict higher gas fractions than the simulation,
because they generally predict less cooling for small haloes
at high redshift. The exceptions are the Somerville and GalICS models, 
which cool most of the halo gas to build up the central galaxies and turn out 
to predict less uncollapsed gas mass for all halo masses.
The simulation shows that cold halo gas contributes a significant fraction
of the halo gas in small haloes. Although the hot halo gas dominates for more massive haloes,
the simulation shows a mild increase in cold halo gas fraction
with redshift for all halo masses. 

All the SAMs predict that the hot fraction in haloes of a given mass increases
with increasing redshift, in contrast to the weak decrease
of the hot fraction with increasing redshift seen in the simulation.
For a given virial mass, haloes at high redshifts have higher densities
than at low redshifts, so they would cool faster. However, the temperature
is higher and the timescale allowing the gas to cool is shorter at 
high redshifts, which can compensate for the density increase and
make cooling slower. The key factor that determines which effect dominates 
is the dependence of the cooling rate on temperature.
If the cooling rate is an increasing function of 
temperature, such as in the Bremsstrahlung regime, the
increase in density with redshift can compensate for the
higher virial temperature and the shorter timescales and can result in
a higher fraction of cooled  halo gas. If the cooling rate drops more rapidly 
with temperature than $T^{-1/2}$, the increase of the virial 
temperature with redshift can overcome the increasing density and 
hence result in less cooling at higher redshifts.
In all the SAMs, we find that even for massive haloes, most of the gas
actually cools at a temperature below $10^5$K, i.e. the bulk of the cooling
occurs in small progenitors. For this temperature range, 
one can expect that the same mass haloes can cool more baryons at late redshifts, which is
consistent with what we find in the SAMs. The simulation, however,
shows the opposite trend. 
The cooling and gas mass assembly is more complex in simulations which
can cause deviations from the trend predicted by the SAMs. The most important 
difference between the SAMs and the simulation is the large
accretion rate of satellite galaxies at high redshift.
These rates are comparable to the central galaxies of the same mass,
which makes gas depletion in high-redshift haloes faster
\citep{Kerevs2009b,Simha2009}. Even though
some satellites can also accrete gas at low redshift, these rates are
typically lower than for central galaxies and, therefore, its effect is 
much weaker. In addition, the UV background can prevent the collapse of
baryons into more massive haloes at late time. At
high redshift the affected mass is below our resolution limit. But
at low-$z$ this mass is close to our resolution limit
\citep{Keres2007}.  This can increase the fraction of uncollapsed
material during the hierarchical build-up of the descendant haloes. Both
of these effects can contribute to produce a flatter, or even reversed
trend of uncollapsed fraction with redshift.

To summarise, gas accretion in SAMs differs significantly from accretion 
in a cosmological hydrodynamical simulation. The intrinsic differences 
between SAM cooling recipes show up as wide dispersions in the SAM results 
at all masses and redshifts, either in the accretion rates or uncollapsed 
gas fractions or both.
Galaxy formation through accretion of cold gas is important
in the simulations but is not modelled at all in SAMs.
SAMs assume that all the gas virialises as soon as it enters
the halo.  The ``rapid'' mode accretion, which is sometimes taken as a proxy
for cold mode accretion generally has lower rates than the simulated cold mode rates.
However, the lack of cold-mode accretion is partially
compensated for by more efficient cooling of the hot halo gas.
This reduces the discrepancy between total accretion rates
in the simulation and the SAMs in small and intermediate mass haloes.
However, a large discrepancy remains at the high-mass end where hot-mode
cooling dominates.

\section{A New Model}\label{sec:new}
%motivation of new model

%SAMs underpredict the cooling rate for
%low-mass haloes but overpredict the rate for massive haloes. 
%Hence, the cooling prescriptions adopted in current SAMs may have missed
%some important aspects of the cooling processes in dark matter haloes. 
As demonstrated in numerical simulations \citep{Kerevs2005, Keres2007,
Kerevs2009}, cold-mode gas accretion plays an important role
in low-mass haloes at all redshifts and in massive haloes at high-redshift.
Although current SAMs include a cooling-radius-based ``rapid'' mode
which may mimic the cold-mode accretion to some
extent,  they do not model the co-existence of cold and hot halo gas 
and the bimodal 
accretion seen in the simulations.  
Motivated by the bimodal accretion
in simulations, we propose a phenomenological model that explicitly
incorporates cold-mode accretion.  We introduce
a cold gas component associated with dark
matter haloes that evolves separately from the hot halo gas.
This gas is assumed to avoid shock heating during infall
and it is not hydrostatically supported.
Furthermore, since cooling is not relevant for this component,
it is assumed to be accreted by the central galaxy in a
free-fall timescale unless the host halo merges into another halo.
We continue to track the hot
component, whose evolution in a halo is assumed to follow
the radiative cooling described in \S \ref{sec:model}. 
SPH simulations \citep{Kerevs2009}  have shown that the majority of
massive haloes develop large cores in their hot gas distribution,
presumably caused by the dynamical heating of recent mergers.
To model this effect we link the
size of the gas core to the mass assembly of the host dark matter
halo, so that the cooling rate for recently formed massive haloes
is reduced.

We describe the  ``bimodal'' nature of cold
gas accretion by the cold and hot halo-gas fractions.
SPH simulations \citep{Kerevs2005, Dekel2006, Keres2007, Ocvirk2008, Kerevs2009, Brooks2009} show that these
fractions depend on both redshift and halo mass.  
In addition, the bimodal accretion has a sharp
transition at a nearly fixed halo mass over a wide redshift range.
Guided by these simulation
results, we model the fraction of hot halo gas for a given virial
mass, $m_{\rm vir}$, as an error function:
\begin{equation}\label{equ:fhot}
p_{\rm hot}(m_{\rm vir}, z) = {f_m(z) \over 2} \left[ 1 + \rm{erf} \left({\log m_{\rm vir} - \log m_{\rm tran} \over \sigma_{\rm logm}}\right)\right],
\end{equation}
where $m_{\rm tran}$ is the transition mass, $\sigma_{\rm logm}$
characterises the sharpness of the transition, and   
$f_m(z)$ describes the maximum hot halo gas fraction for a halo
with $m_{\rm vir}\gg  m_{\rm tran}$ at redshift $z$.
Simulations show that the accretion in massive haloes is dominated
by the hot mode at low redshifts but the contribution of the
cold mode increases with increasing redshift.
We therefore model $f_m$ as a decreasing function of redshift,
\begin{equation}
f_m(z) = f_{\rm c} e^{-({z \over z_{\rm tran}})^2} + \left(1-f_{\rm c}\right),
\end{equation}
where $z_{\rm tran}$ is a transition redshift and 
$f_{\rm c}$ is the fraction of the cold halo gas in very massive haloes
at $z\gg z_{\rm tran}$. 
Our model for hot gas fraction, therefore, is specified by three free
parameters.
We fix the values of these parameters
by matching the model predictions with the accretion rate and 
halo gas fraction (in both the cold and hot modes) obtained
from the simulations. This gives $m_{\rm tran}\approx 10^{11.4}\msun$,
$\sigma_{\rm logm}\approx 0.4$,  $z_{\rm tran}=4$ and $f_{\rm c}=0.1$.

This parametrisation serves only as an illustration that bimodal accretion
matching a particular simulation
can be modelled and straightforwardly implemented in SAMs.
A general treatment must be based on a physical model for
the dependence of the cold and hot gas fraction on the
resolution, metal cooling, halo assembly history, and
feedback processes. Each of these effects can change the relative
distribution of the hot and cold components. 

We apply the above model to merger trees to predict the
evolution of the different gas components in a halo by
calculating the hot fraction, $p_{\rm hot}$, for a halo of mass
$\mvir$ at redshift $z$. This fraction of the baryonic mass in the
halo is assumed to be at the virial
temperature, while the rest is assumed to be in the cold component.
To make predictions for the accretion rate of the cold
component onto the central galaxy,  we assume that the accretion
rate is half of the total mass of cold halo gas, $m_{\rm ch}$, divided by
the free-fall  timescale,
\begin{equation}
\dot{m}_{\rm cold}={m_{\rm ch} \over 2\tau_{\rm dyn}} 
= {m_{\rm ch} V_{\rm c} \over 2 \rvir}.
\end{equation}
Although the cold gas is likely to infall on a timescale close to the
free-fall time, we only allow half of the total cold mass to collapse
onto the central galaxy in a free-fall time to mimic several effects 
that are not included in the current model: some of the material can be
accreted by satellites, a fraction of the currently cold
gas can get heated to the virial temperature, and the angular momentum of
the infalling gas must be removed before it can be accreted by the
central galaxy.

We redistribute the hot gas only at the birth time of a
halo. Following the Cole model, a halo is born when its 
mass is doubled. Motivated
by the simulation results described above, we assume a cored profile
for the hot halo gas, with a core radius, $r_c$ being half of 
the scale radius of the halo profile, $r_s$ \citep[assumed to have an 
NFW form, see][]{Navarro1997}, i.e. $r_c=0.5 r_s$.
$N$-body simulations have shown that the halo
concentration parameter, $c=\rvir/r_s$, is closely related to the
formation time of a halo \citep{Bullock2001a, Wechsler2002, Zhao2003a,
  Zhao2003, Zhao2009}.  We adopt the following simple form for this relation:
\begin{equation}
c=3  \left({1+z_c \over 1+z}\right),
\end{equation}
where $z$ is the redshift at selection
and $z_c$ is the redshift at the birth time.
This allows us to estimate $r_s$, and hence $r_c$, for
any halo at any redshift. Using this prescription, we find that
$10^{11}\msun$ haloes at the present time
typically have a concentration of about 15,
while haloes of $10^{15}\msun$ have a concentration of about 5.

From the hot gas profile, we calculate the cooling radius,
$r_{\rm cool}$, and the free-fall radius, $r_{\rm ff}$, and assume
that only the hot gas within both of these two radii can
cool and accrete onto the central galaxy.
Thus, the effective radius for cooling is
$r_{\rm min}(t)=\min[r_{\rm cool}, r_{\rm ff}, \rvir]$.
The cooling rate of the hot mode is assumed to be
equal to the mass of hot gas between spherical shells
of radii given by the values of $r_{\rm min}$ at the current
time and at one time step earlier.  At each generation,
the newly accreted hot gas is included by
changing the normalisation of the hot gas profile at each time step,
following the Somerville model.

%summary
Figures \ref{fig:mcr} and \ref{fig:mcm} show that
the new model predicts higher
cooling rates for low-mass haloes, but lower cooling rates for
massive ones compared to the other models. In the new model, 
low-mass haloes accrete gas mostly through
cold mode, and massive haloes cool gas from their hot haloes at a
reduced rate because they typically form late and hence maintain a big
core in  their hot gas distribution.
Figure \ref{fig:mcf} shows the fraction of cold accretion
predicted by the new model, which is the ratio of the accretion
rate of the cold halo gas to the total accretion rate. 
Clearly, in the new model
the cold mode dominates more and the transition from cold
to hot modes occurs at lower redshifts. 

Figures \ref{fig:simu-sam} and \ref{fig:simu-sam-hmf} compare the new
model with the simulation.
The new model nicely reproduces the accretion rates for
both the cold and hot modes for all halo masses over a large range of
redshift and the predicted mass fractions in the hot and cold halo gas
are in rough agreement the simulation results.
However, as in the other models, the new model also predicts a weak
increasing trend for the hot fraction 
with increasing redshift in contrast to the weak decreasing 
trend seen in the simulation (see Section \ref{subsec:simu}). 

\begin{figure}
\epsfig{file=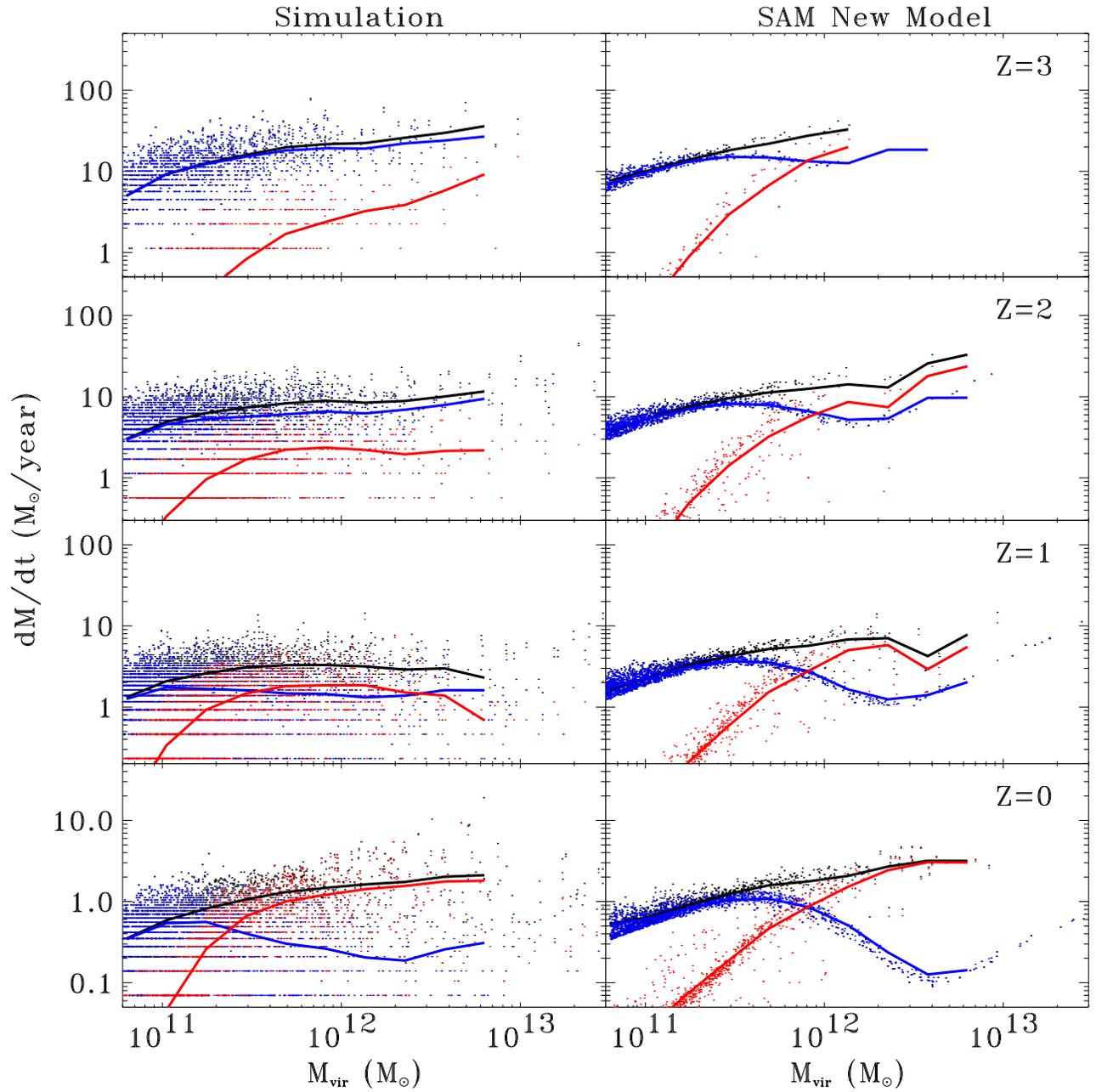}
\caption{\large 
The total rate of accretion (black) and  the ``cold'' (blue) and ``hot'' (red) mode
contribution for individual central galaxies in haloes in the 
simulation (left column) and the new model (right column) at four
redshifts, which are noted in the upper-right corner of each right
panel. The red, blue and black lines show the mean hot, cold and total accretion rates. 
}
\label{fig:simu-sam-sct}
\end{figure}

To compare the new model with the simulation results in more detail,
we show the predicted and
simulated accretion rates in both the cold and hot modes for
individual haloes at four different redshifts, $z=0,1,2$ and 3 in Figure \ref{fig:simu-sam-sct}.
The new model matches the simulation results much better
than any other model considered in this paper. In particular,   
the model shows the cold--hot-mode transition at
$\sim 7\times10^{11}\msun$. 
Although the transition mass has a complex redshift dependence 
in the simulation, the model captures the characteristic mass scale. 
However, the predicted scatter in the accretion rate for a given
halo mass is much smaller than that seen in the simulation.
For a given merger history, the model ignores
all stochasticity that may arise from different merger orbits
and from the interactions between different mass components
of a halo. The large scatter seen in the simulations suggests 
that these effects may have a significant impact on the
gas accretion rates in individual haloes.     

\begin{figure}
\epsfig{file=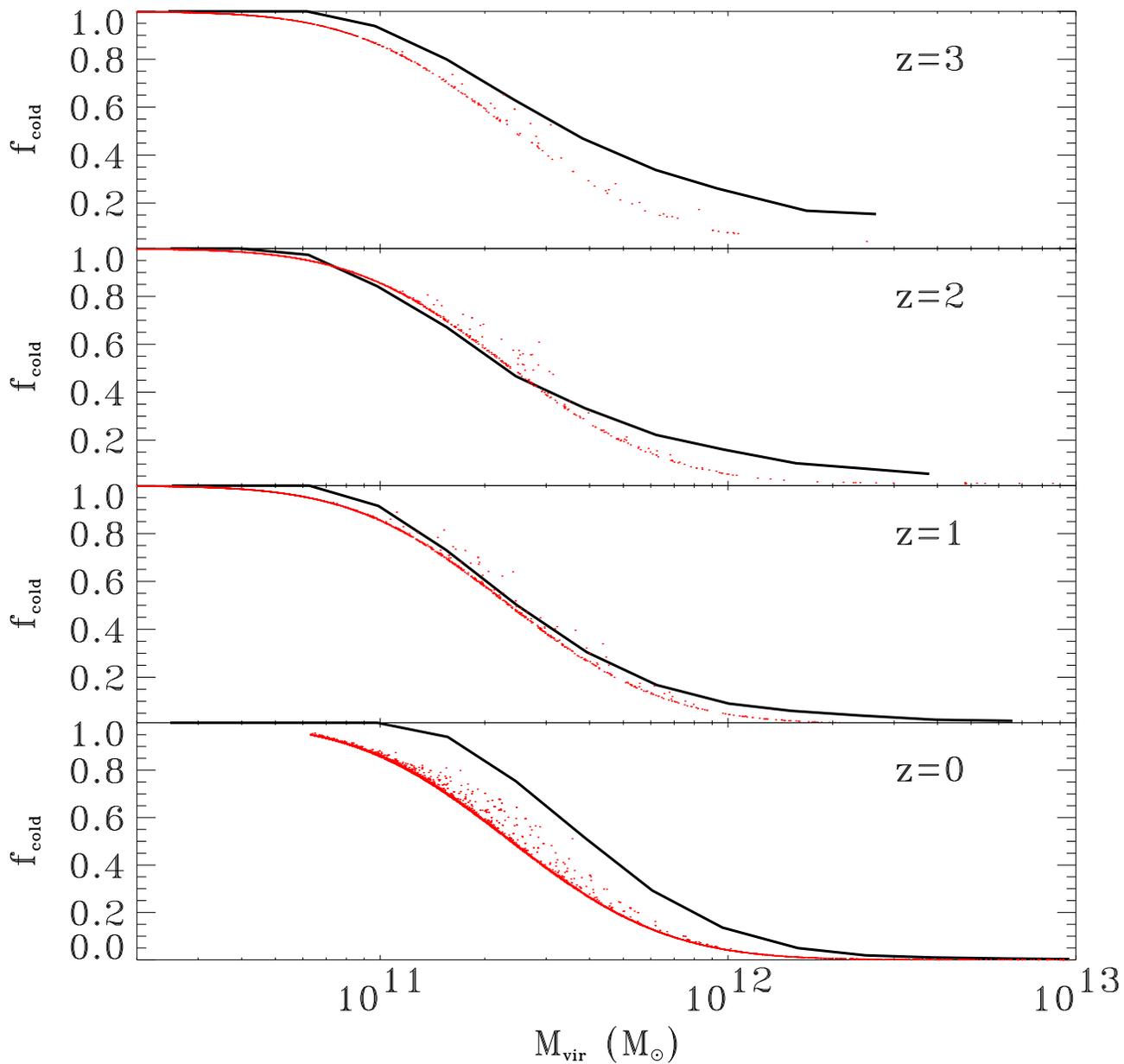}
\caption{\large The fraction of cold halo gas in a halo of given mass at four redshifts. Lines
show the median of the SPH simulation results, and dots show a randomly selected sample of haloes
from the new model SAM.
\label{lastpage}
}\label{fig:simu-sam-fcold}
\end{figure}

Figure \ref{fig:simu-sam-fcold} compares the fraction of cold
halo gas for a given halo mass predicted by the new model 
with that from the simulation.
The model reproduces the overall trend seen in the
simulation but under-predicts the cold fraction at $z=0$.
The discrepancy is a consequence of the temperature
criterion used to select cold gas in the simulation. 
Following \cite{Kerevs2005}, we demarcate the
cold and hot component by $T=2.5\times10^5 \rm K$, but this is
comparable to the virial temperature for haloes around the
transition from mostly cold to mostly hot gas at $z\sim 0$. Therefore, gas
in the halo outskirts that is typically slightly colder will be
added into the cold component even if a fraction of such gas was shock
heated to $\sim T_{\rm vir}$. At high
redshift this is not an issue since the $T_{\rm vir}$ of these
haloes is much higher.

The new model separately tracks 
the cold and hot phases of the halo gas.
As we have demonstrated in this section, the two-phase model not only
reproduces
the bimodal accretion seen in simulations of a CDM universe, but also 
can be tuned to match the simulation results. 
However, our phenomenological model does not include the
physical processes that determine the division between hot and cold
halo gas.  Therefore, the model parameters presented in this paper
apply only when the feedback does not strongly affect the cold-mode
accretion we tune the parameters to match a simulation
without strong feedback. 
Feedback and preheating mechanisms may
alter the fraction and evolutionary track of the cold halo gas.
Further work is needed to better understand the physics of cold-mode 
accretion \citep[e.g.][]{Oppenheimer2010}, such as
the effect of feedback on cold-mode accretion
and the interaction between the cold and hot halo gas phases.
While our model is purely phenomenological, future physical models need 
to properly model the change in the accretion rates of cold and hot gas as a
function of radius in a halo
\citep[e.g.][]{Faucher-Giguere2011, vandeVoort2010}.

\section{Discussion and Conclusions}\label{sec:dis}

The process of radiative cooling is fundamentally important for modelling 
galaxy formation.  This process is often taken
for granted, but, as we have shown in this paper, the predictions of 
existing cooling prescriptions in semi-analytic models (SAMs)
differ significantly.
We implemented the numerical recipes used in six published SAM codes,
and compared their predictions with each other and hydrodynamic simulations 
for idealised spherical, smooth-accretion halos and for realistic 
merging halos in a cosmological context. 
Our tests examined the central galaxy accretion rates, the total cooled baryonic
masses, and the fraction of baryons remaining in a hot halo.
We focused on a range of halo masses relevant for galaxy formation
and found large variance between different SAMs themselves and between SAMs and 
the cosmological SPH simulation.

These comparisons 
are useful for understanding the interdependence of our phenomenological
descriptions
and illustrate the need for a more reliable cooling model.
For example, the cooling algorithms control gas accretion and gas
phase fractions, these
will affect star 
formation algorithms, star formation driven feedback, and AGN heating. 
The discrepancies in the models suggest the possibility of errors in 
both the physical interpretation and the predictions from SAMs. Furthermore, 
incomplete or incorrect modelling of the halo gas
component will directly affect predictions for properties of halo
gas and prospects for direct detection of different gas components. 

Although our paper compares the largest collection of 
SAMs to a hydrodynamic simulation to date, there
have been several attempts to compare a single SAM with a SPH simulation
\citep[see][]{Benson2001c, Yoshida2002, Helly2003, Cattaneo2007}.
All of these found good agreement between the ``stripped-down'' SAMs and the simulations. 
However,  \citet{Springel2002} have shown that the SPH implementation 
affects gas cooling rates, and
as we have seen, subtle choices in SAM implementation may result in
dramatic changes in the gas accretion rate.  In some
cases, we speculate that the reported consensus may be the result
of insensitive tests; in
others, it may reflect an overzealous match between SPH implementation and
the choice of SAM cooling algorithm. Indeed, 
a recent SAM-simulation comparison \citep{Saro2010}  
also shows that the accretion rate of the central galaxies in massive  
haloes from SAM and SPH simulations differ significantly. 

Comparisons between competing SAMs are less numerous in the literature, 
but a recent paper by \citet{DeLucia2010} compares three models, the 
``Durham'' (effectively Cole with a core), ``Munich'' (effectively
Croton), and MORGANA models \citep{Monaco2007}.  These comparisons are
complementary
to our tests.  Their results are
similar to ours, e.g. the lower cooling rates in the Durham
model in the ``rapid'' mode at early times and for massive haloes 
at late times.  Moreover, our extended comparison to six
SAMs reveals an even larger dispersion among models
than that found by \citet{DeLucia2010}.

In the absence of strong feedback, most cooled gas in the universe
comes from cold-mode accretion where the gas 
infalls onto galaxies without heating to the virial temperature.  
This process, therefore, should be central to the cooling models in SAMs.  
Currently, the ``rapid'' mode in SAMs is a proxy for this mechanism.
However, our study highlights a large variance between the
predictions for different SAM implementations.  In part, this owes to different
cooling radius calculations and timescales for ``rapid'' mode gas accretion.
We also find that SPH simulations predict larger cold-mode
accretion rates than the SAM ``rapid'' mode for some or 
all halo and redshift regimes, depending on the SAM.
In addition, the cooled gas mass in the simulation 
has an order 30\% dispersion, which is much less than that in 
the ``rapid'' mode accretion rates predicted by the SAMs.
Even in the regimes where the total accretion rates are
comparable between the SAMs and simulation, the properties and
detectability of the accreting halo gas may very much depend on
the accretion mode \citep[e.g.][]{Fardal2001, Dijkstra2009, 
Goerdt2010, Faucher-Giguere2010}. 
Furthermore the angular momentum of the
infalling material can differ between the accretion modes
\citep{Kerevs2005, Benson2010b}. Different physics of gas accretion in
the cold and hot mode clearly needs to be captured properly in a model in
order to understand the formation and evolution of galaxies. 

To make progress, we propose a new cooling algorithm that 
introduces the cold-mode gas explicitly.  The new model
better matches the accretion rates 
and both the hot and halo gas mass fractions from the simulations.
In addition, the explicit cold-mode channel
can handle effects like aspherical and 
simultaneous cold/hot
accretion.
Other recent work on including cold-mode accretion explicitly in SAMs include
\citet{Kang2010}, who used a simple sharp transition from cold to hot haloes
depending on halo mass, and \citet{Benson2010b}, who more
closely resembles our gradual transition. \citet{Benson2010b} 
find that strong SN feedback can always 
heat the ejected gas to high temperature in the halo
and hence argue that the substitution of an explicit cold mode for
their ``rapid'' mode made a negligible difference to their results. 
\citet{Kang2010} demonstrate
that changes in the cold mode are significant for high-redshift, high-mass 
galaxies, and this impacts the prediction of global star formation rates.
The discrepancy between these results suggest that effect of the cold-mode accretion depends on 
the implementation of feedback \citep{DeLucia2010}. However, the interaction between feedback 
and the infalling cold gas is not yet clear. High redshift cold-mode
accretion proceeds in filamentary streams of cold gas which are hard
to destroy by outflowing material, especially because the outflows
will tend to expand into the low density regions in the halo. In fact,
\citep{Oppenheimer2010} show that the high redshift cold-mode accretion 
seems to be insensitive to the gas ejected with milder velocities.
Observations that probe the properties of the halo gas could
provide constraints on the feedback models and their effect on the
cold halo gas. Moreover, it has been pointed out that ``preventive'' feedback,
e.g. radio mode AGN, is unlikely to affect the gas accretion in cold
mode with the same efficiency as it would affect gas in a hot halo
\citep{Kerevs2009b}. Therefore, we would argue that a more accurate
treatment for gas accretion in SAMs is crucial to better understand
how feedback works in galaxy formation.  

This paper demonstrates that cooling processes in current semi-analytic models 
produce predictions that are at variance with each other and
numerical simulations.   We advocate a careful normalisation of the cooling 
models using a combination of numerical simulations and analytic arguments 
to improve the prescriptions and observational comparisons to 
constrain the cooling channels.
For example, star-formation timescales are expected 
to be short at high redshift, so the instantaneous star formation rate should provide 
a lower limit on the ongoing gas accretion. Such observations
\citep[e.g.][]{ForsterSchreiber2009} are already providing challenging
constraints for cooling models in SAMs \citep[see][]{Khochfar2009}.  
To derive the constraints on gas accretion
histories, we have implemented
SAMs as a Bayesian inference problem and have demonstrated it leads to genuine
constraints on model parameters and on model prescriptions themselves \citep{Lu2010a}.  
For example, the Bayesian approach will allow us to assess the observational support of our new cold-mode
accretion prescription relative to the existing cooling prescription 
using
Bayes factors.  Alternatively, gas accretion histories may be
constrained with one or more observational data sets with non-parametric models
\citep{Neistein2010}.
Most likely, a variety of approaches will be necessary 
to improve our treatment of gas cooling, a key step 
in understanding how galaxies form and evolve.

%%%%%%%%%%%%%%%
% Acknowledgments
%%%%%%%%%%%%%%%
\section{Acknowledgments}

We thank Andrew Benson, Darren Croton,
and Martin White for useful discussion.
DK acknowledges supports from Harvard University funds and from NASA through Hubble
Fellowship grant HSTHF-51276.01-A.
This work was supported in part by NSF IIS Program through award 0611948
and by NASA AISR Program through award NNG06GF25G.
HJM acknowledges the support of NSF 0908334. 

%%%%%%%%%%%%%%%
% Bibliography
%%%%%%%%%%%%%%%

%\begin{thebibliography}{}
\bibliography{/Users/luyu/references/general}
%\end{thebibliography}

\newpage

\newpage

\newpage

\newpage

\newpage

\newpage

%%%%%%%%%%%%%%%%%%%%%%%%%%%%%%%%%%%%%%%%%%%%%%%%%%%%%%%%%%%%%%
\newpage

%%%%%%%%%%%%%%%%%%%%%%%%%%%%%%%%%%%%%%%%%%%%%%%%%%%%%%%%%%%%%%

\newpage

%%%%%%%%%%%%%%%%%%%%%%%%%%%%%%%%%%%%%%%%%%%%%%%%%%%%%%%%%%%%%%

\newpage
%%%%%%%%%%%%%%%%%%%%%%%%%%%%%%%%%%%%%%%%%%%%%%%%%%%%%%%%%%%%%%

\newpage

%%%%%%%%%%%%%%%%%%%%%%%%%%%%%%%%%%%%%%%%%%%%%%%%%%%%%%%%%%%%%%

\end{document}